\documentclass[fleqn,usenatbib]{mnras}

\usepackage{newtxtext,newtxmath}

\usepackage[T1]{fontenc}

\DeclareRobustCommand{\VAN}[3]{#2}
\let\VANthebibliography\thebibliography
\def\thebibliography{\DeclareRobustCommand{\VAN}[3]{##3}\VANthebibliography}

\usepackage{graphicx}	
\usepackage{amsmath}	

\def\ha{{\rm\,H$\alpha$}}

\def\pab{{\rm\,Pa$\beta$}}

\def\nii{{\rm\,[N{\sc ii}]}}

\title[Enhanced star formation in low-mass galaxies]{Spider-Webb: enhanced star formation in low-mass galaxies within the Spiderweb protocluster revealed by JWST \pab\ narrow-band imaging}
\author[K. Daikuhara et al.]{Kazuki Daikuhara,$^{1,2}$\thanks{E-mail: daikuhara@ir.isas.jaxa.jp}
Tadayuki Kodama,$^{2}$
Yusei Koyama,$^{3}$
Rhythm Shimakawa,$^{4}$
Helmut Dannerbauer,$^{5,6}$
\newauthor
Jose Manuel Pérez-Martínez,$^{5,6}$
Pablo G. Pérez-González,$^{7}$
Mariko Kubo,$^{2,8}$
Eduardo Ibar,$^{9,10}$
\newauthor
Philip N. Best,$^{11}$
Abdurrahman Naufal,$^{3,12}$
Yuheng Zhang,$^{5,13,14}$
Ronaldo Laishram$^{2,3}$
\\
$^{1}$Institute of Space and Astronautical Science, Japan Aerospace Exploration Agency, 3-1-1, Yoshinodai, Chuou-ku, Sagamihara, Kanagawa 252-5210, Japan\\
$^{2}$Astronomical Institute, Tohoku University, 6-3, Aramaki, Aoba, Sendai, Miyagi, 980-8578, Japan\\
$^{3}$National Astronomical Observatory of Japan (NAOJ), National Institutes of Natural Sciences, 2-21-1 Osawa, Mitaka, Tokyo 181-8588, Japan\\
$^{4}$Waseda Institute for Advanced Study (WIAS), Waseda University, 1-21-1, Nishi-Waseda, Shinjuku, Tokyo 169-0051, Japan\\
$^{5}$Instituto de Astrofísica de Canarias (IAC), E-38205, La Laguna, Tenerife, Spain.\\
$^{6}$Departamento Astrofísica, Universidad de La Laguna, E-38206 La Laguna, Tenerife, Spain\\
$^{7}$Centro de Astrobiología (CAB), CSIC-INTA, Ctra. de Ajalvir km 4, Torrejón de Ardoz, E-28850, Madrid, Spain\\
$^{8}$School of Science, Kwansei Gakuin University, Sanda, Hyogo 669-1337, Japan\\
$^{9}$Instituto de F\'isica y Astronom\'ia, Universidad de Valpara\'iso, Avda. Gran Breta\~na 1111, Valpara\'iso, Chile\\
$^{10}$Millennium Nucleus for Galaxies \\
$^{11}$Institute for Astronomy, University of Edinburgh, Royal Observatory, Blackford Hill, Edinburgh EH9 3HJ, UK\\
$^{12}$Graduate University for Advanced Studies (SOKENDAI), 2-21-1 Osawa, Mitaka, Tokyo 181-8588, Japan\\
$^{13}$School of Astronomy and Space Science, Nanjing University, Nanjing, Jiangsu 210093, China\\
$^{14}$Key Laboratory of Modern Astronomy and Astrophysics (Nanjing University), Ministry of Education, Nanjing 210093, China
}

\date{Accepted 2026 May 01. Received 2026 April 29; in original form 2026 February 13}

\pubyear{2026}

\begin{document}
\label{firstpage}
\pagerange{\pageref{firstpage}--\pageref{lastpage}}
\maketitle

\begin{abstract}
Understanding the role of the environment in galaxy evolution is key to revealing the physical processes that regulate galaxy growth. 
We study the star formation activity of \pab\ emitters (PBEs) in the Spiderweb protocluster at $z=2.16$ using \textit{James Webb Space Telescope}/NIRCam narrow-band imaging.
To investigate the environmental dependence of star formation, we derive star formation rates (SFRs) from the \pab\ emission line and compare SFRs in the Spiderweb protocluster with those in the field.
Our main finding is that low-mass PBEs ($M_\star < 10^9\,M_\odot$) in the Spiderweb protocluster exhibit an enhancement in star formation compared to their field counterparts.  
This excess persists even without applying dust-attenuation corrections, indicating that enhanced star formation in the protocluster is robust regardless of whether a dust correction is applied.
In contrast, intermediate- and high-mass PBEs ($M_\star > 10^9\,M_\odot$) show no significant deviation from the field, revealing a strong mass dependence in the environmental effects on star formation.  
No clear spatial concentration toward the cluster core of starbursting low-mass galaxies within the protocluster is seen, suggesting that their enhancement is not restricted to the cluster core.  
We suggest that starbursts in low-mass galaxies are facilitated by environmental processes such as galaxy mergers/interactions, and/or efficient gas supply.  
While the enhancement at the low-mass end is consistent with trends reported for other protoclusters at similar redshifts, the behaviour of star formation at intermediate masses ($10^{9} < M_\star/M_\odot < 10^{10}$) is not uniform across protoclusters.  
Our \pab-based results in the Spiderweb protocluster indicate that star-formation enhancement at cosmic noon depends on both mass and the dynamical state of the protocluster.
\end{abstract}




\section{Introduction}
In the $\Lambda$CDM cosmological framework, small dark matter haloes form first and subsequently merge and accrete to become more massive systems \citep{PressSchechter1974, WhiteRees1978, Blumenthal1984, Lacey1993}.
Galaxies evolve with their host haloes, building stellar mass through in-situ star formation and mergers associated with halo mass accretion histories \citep{Behroozi2013, Wechsler2018}.
Thus, low-mass galaxies at early epochs represent the fundamental building blocks of present-day massive systems.
Low-mass galaxies are also particularly sensitive to feedback processes, making them crucial laboratories for studying how baryonic physics shapes galaxy evolution \citep[e.g.,][]{Muratov2015, Pandya2021, Mitchell2020, Pillepich2018}.

The importance of environment in shaping galaxy properties has been recognized since early studies of galaxy clusters, which revealed strong correlations among galaxy morphology, star formation, and local density \citep[e.g.,][]{Dressler1980,Boselli2006,Boselli2014,Overzier2016,Cortese2021}.
In the local Universe, the star formation activity of low-mass galaxies is strongly affected by their surrounding environments in such a way that the quenching is more proceeded in dense regions \citep[e.g.,][]{Peng2010, Geha2012}.
However, it remains unclear whether such environmental effects were already established during the early stages of galaxy assembly.

The Butcher–Oemler effect \citep{Butcher1984} shows that the fraction of blue star-forming galaxies in galaxy clusters increases with redshift.
This trend demonstrates that the galaxy population in dense environments evolves strongly with cosmic time. 
Following this pioneering work, numerous studies  of galaxy populations in clusters have been conducted to know how galaxy populations in dense environments evolve over cosmic time, and these studies are consistently finding that the fraction of blue, star-forming galaxies in clusters increases toward higher redshifts \citep[e.g.,][]{Poggianti1999,Kodama2001,Nantais2017,vanderBurg2020}.
These results collectively indicate that galaxy populations in dense environments have undergone substantial evolution across cosmic time.
In parallel, the star formation activity of galaxies in general evolves strongly with redshift, as reflected in the change of normalization of the star-forming main sequence \citep[hereafter MS; ][]{Noeske2007}, which has been extensively studied \citep[see][]{Brinchmann2004, Daddi2007, Elbaz2007, Salim2007, Magdis2010, Rodighiero2011, Wuyts2011, Whitaker2012, Whitaker2014, Heinis2014, Speagle2014, Schreiber2015, Shivaei2015, Tomczak2016, Santini2017, Iyer2018, Leslie2020, Leja2022, Daddi2022, Popesso2023,Koprowski2024}.
Examining how star-forming galaxies populate the MS across different environments and cosmic epochs therefore provides a powerful framework for quantifying environmental effects on galaxy evolution.

Whether the star-forming MS depends on environment has been actively debated. 
At low redshift ($z<1$), star-forming cluster galaxies follow essentially the same MS as field galaxies \citep[e.g.,][]{Peng2010, Wijesinghe2012}, although some works report a modest suppression of SFRs in dense environments \citep[e.g.,][]{Vulcani2010}. 
During the epoch of peak cosmic star formation ($1<z<3$, “cosmic noon”), some studies conclude that the star-formation activity shows little or no environmental dependence in star-forming galaxies \citep[e.g.,][]{Koyama2014,Darvish2016,Jose2023}, while mild suppression \citep[e.g.,][]{Old2020} or even enhanced star formation of low-mass galaxies in a protocluster \citep[][]{Daikuhara2024} has also been reported. 
However, star-forming low-mass galaxies at cosmic noon remain poorly explored due to observational challenges. 

Theoretical studies predict that in dense environments such as protoclusters around cosmic noon, the mode of gas supply is expected to undergo significant changes \citep{Dekel2006}.
In haloes above and below the critical shock-heating mass ($M_{\rm shock} \sim 10^{12} M_\odot$), stable shocks can develop and sustain hot gaseous haloes.
However, even in such environments, dense cold streams along the cosmic web can penetrate through the hot medium and efficiently deliver gas to the central regions.
With increasing halo mass and cosmic time, however, these streams are gradually suppressed within the hot halo and fail to reach the central regions, leading to a decline in cold gas supply and star formation activity \citep{Keres2005, Dekel2009, Umehata2019, Daddi2021}.

Cosmic noon ($z\sim2$–3) marks a pivotal epoch when galaxies undergo the transition from being predominantly sustained by cold stream accretion to a regime increasingly dominated by shock-heated haloes.
Observational studies at $z\sim2$–3 support this picture, indicating that variations in gas supply can strongly influence both enhanced star formation activity and chemical enrichment histories in dense environments \citep[e.g.,][]{Daddi2021, Jose2023, Jose2024,Daikuhara2024}.

We aim to understand how galaxy environments control star formation during the epoch of cosmic noon, in order to reveal the physical processes that drive galaxy evolution.
Exploring low-mass and dusty star-forming galaxies in protoclusters at $z > 2$ is particularly important, because these populations remain difficult to identify in previous surveys \citep{Hayashi2016,Daikuhara2024}.
Our study is designed to probe such galaxies and to examine how their star formation depends on environment.

As part of the Spider-Webb project (JWST Cycle 1 GO 1572 programme; PIs: H. Dannerbauer and Y. Koyama), we build a sample of narrow-band (NB) selected \pab\ emitters (PBEs) in and around the Spiderweb protocluster \citep{Shimakawa2024a}.
Using deep NB imaging targeting the \pab\ emission line, we identify star-forming galaxies with a dust-insensitive tracer down to low stellar masses, enabling a robust census of star formation across a wide mass range within a single protocluster at the epoch of cosmic noon.
We also apply the same analysis methods to \textit{James Webb Space Telescope} (JWST) Emission Line Survey data \citep{Duncan2025,Pirie2025} at similar redshifts, enabling a fair comparison between protocluster and field galaxies.

In this paper, we focus on the entire PBE sample, while complementary X-ray/AGN-oriented and dust-attenuation analyses of massive \ha\ emitters are presented in \citet{Shimakawa2024b} and \citet{Jose2024a}, respectively.
Section 2 introduces the properties of the Spiderweb protocluster, while Section 3 describes the observational data and analysis methods.
Our results are presented in Section 4, followed by a discussion in Section 5, and we summarize our conclusions in Section 6.

Throughout this paper, we adopt a flat $\Lambda$CDM cosmology with $H_{0} = 70\ \mathrm{km\ s^{-1}\ Mpc^{-1}}$, $\Omega_{\mathrm{M}} = 0.3$, and $\Omega_{\Lambda} = 0.7$.
All magnitudes are given in the AB system \citep{Oke1983}, and stellar masses and star formation rates are derived assuming a Chabrier initial mass function (IMF; \citealp{Chabrier2003}).

\section{Spiderweb protocluster}\label{sec:data}

The Spiderweb protocluster at $z = 2.2$ is a spectacular example of a high-$z$ overdense structure associated with a central radio galaxy, PKS1138–262 at $z = 2.16$ \citep[Spiderweb: ][]{Miley2006}. 
This structure was also originally discovered as an overdensity of LAEs \citep{Kurk2000} around the Spiderweb radio galaxy \citep{Roettgering1994,Pentericci1997}.
The structure of the Spiderweb protocluster was also characterized by \ha\ NB \citep{Hatch2011,koyama2013,Shimakawa2018pks,Daikuhara2024}, \pab\ NB \citep{Shimakawa2024a,Shimakawa2024b,Jose2024a,Laishram2026}, spectroscopy \citep{Tanaka2013,Shimakawa2015,Jose2023,Naufal2024}, red sequence galaxies \citep{Kodama2007,Zirm2008,Tanaka2010,Tanaka2013}, submillimetre galaxies \citep{Dannerbauer2014}, and (sub-) millimetre observations to trace the dust content, gas reservoir and cluster gas \citep{Dannerbauer2017,Emonts2018,Tadaki2019,Jin2021,DiMascolo2023,Zhang2024,Chen2024,Zhang2026}, and X-ray follow-up observations \citep{Pentericci2002,Croft2005,Tozzi2022}.

The physical association of this structure was confirmed by near-infrared (NIR) spectroscopy performed with Subaru/MOIRCS \citep{Shimakawa2014}, CO observation \citep{Jin2021,Jose2025}, VLT-KMOS spectroscopy \citep{Jose2023} and ALMA 1.2 mm mapping \citep{Zhang2024}, and the dynamical mass of this cluster is estimated to be $M_{\mathrm{cl}} \sim 1.7 \times 10^{14}\ \mathrm{M_{\odot}}$.
Recently, the thermal Sunyaev–Zeldovich effect was confirmed in the Spiderweb protocluster by \cite{DiMascolo2023}, suggesting hot cluster gas is building in the protocluster core.
They estimated that the Spiderweb protocluster has a halo with $M_{500}=3.46 \times 10^{13}\ \mathrm{M_{\odot}}$ and $r_{500} = 288.9$ kpc.
Compared with the field galaxies, the Spiderweb protocluster had a higher fraction of mergers \citep[$\times  1.5$;][]{Naufal2023}, quiescent galaxies \citep[$\times  3$;][]{Naufal2024}, number density of dusty star-forming galaxies \citep{Zhang2024}, number density of red sequence galaxies \citep{Kodama2013}, metallicity \citep{Jose2023}, and AGNs \citep{Tozzi2022}.

\section{Method}
\subsection{Data set}
\label{subsec:dataset}
We utilize the catalogue of PBEs constructed by \citet{Shimakawa2024a}.
That study conducted JWST narrowband (NB) imaging (Cycle 1, GO1572; PIs: H. Dannerbauer and Y. Koyama) with NIRCam, using the F405N, F410M, F115W, and F182M filters.
The raw data are available at the Barbara A. Mikulski Archive for Space Telescopes (doi:10.17909/vx25-q902).
The NIRCam F405N filter ($\lambda_{\mathrm{c}} = 4.055\ \mu$m, FWHM $= 0.046\ \mu$m) captures the \pab\ line at the protocluster redshift of $z = 2.16$.
A detailed description of the data analysis is provided in \citet{Shimakawa2024a}.
Emitters were selected on the basis of the F410M$-$F405N colour excess and a colour–colour diagram.
The typical flux limit is $\sim 2 \times 10^{-18}\ \mathrm{erg\, s^{-1}\, cm^{-2}}$, corresponding to a rest-frame equivalent width (EW) limit of $\sim 20$ \AA.
A total of 41 PBEs that satisfy the selection criteria were identified.
We exclude the radio galaxy PKS1138 from our analysis owing to the complexity of its spectral energy distribution (SED).
Since this study primarily focuses on low-mass galaxies, this exclusion does not affect our conclusions.

We construct a multi-band catalogue of the PBEs using the JWST filters F405N, F410M, F115W, and F182M \citep[][]{Shimakawa2024a} and HST ACS/WFC filter F475W, F814W \citep[PI: H. Ford, proposal ID 10327;][]{Miley2006} and HST/WFC3 F160W \citep[PI: Y. Koyama, proposal ID 17117;][]{Naufal2024}.
To ensure consistency across datasets with different spatial resolutions, we perform point spread function (PSF) matching by degrading the JWST images to match the HST F160W full width at half maximum (FWHM), using Gaussian kernels.
The final PSF size is $0^{\prime\prime}.19$, and the pixel scale is $0^{\prime\prime}.06$ per pixel.
This procedure ensures that photometry is directly comparable across bands and instruments.

Source extraction is performed on the median-combined image of the four JWST bands (F115W + F182M + F410M + F405N) with SExtractor \citep[version 2.25.2;][]{Bertin1996}.
Photometry is carried out in double-image mode with Kron apertures \citep{Bertin1996}.
To obtain reliable stellar mass estimates, we adopt Kron magnitudes rather than fixed apertures. 
Kron apertures follow the actual light distribution of each galaxy and typically enclose about 90\% of the total flux, which mitigates systematic biases introduced by fixed-aperture photometry.

Estimation of photometric uncertainties is crucial for robust measurements of faint emission-line sources.
To account for spatial variations in background noise, we evaluate the noise level from the weight image following the procedure of \citet{Shimakawa2024a}.
Background noise ($\sigma$) is estimated through random aperture photometry using the {\tt photutils} package \citep{Photutils}, following methods described in \citet{Bunker1995, Shimakawa2018pks, Shimakawa2024a, Daikuhara2024}.
By examining the error distribution as a function of aperture size and weight, we compute photometric errors appropriate for any chosen aperture \citep[see ][ for details]{Daikuhara2025b}.
This approach properly accounts for spatial variations in the background and provides reliable error estimates across the mosaic.

\subsection{Spectral energy distribution}
\label{subsec:sed}
We perform SED fitting to derive stellar masses ($M_{\star}$) and dust attenuation using the {\it CIGALE} code \citep[Code Investigating GALaxy Emission, version 2025.0;][]{Burgarella2005,Noll2009,Boquien2019}.
The stellar population synthesis models are based on \citet{BruzualCharlot2003}, adopting a \citet{Chabrier2003} initial mass function (IMF) and an exponentially declining and delayed star formation history of the form SFR $\propto t \exp(-t/\tau)$, with e-folding timescales ($\tau$) ranging from $10^9$ to $10^{10}$ yr. 
We also include nebular emission model using the CIGALE nebula module.
The stellar attenuation ($A_V$) is allowed to vary from 0 to 3.0 mag, and we assume $E(B-V){\mathrm{gas}} = E(B-V){\mathrm{stellar}}$.
The metallicity of the stellar population is fixed at $Z = 0.004$, following previous MAHALO-deep studies \citep{Daikuhara2024}.
A complete list of parameters is provided in Table~\ref{table:cigale}.

It should be noted that stellar masses may be overestimated if an AGN is present, as AGN light contributions are not accounted for in the fitting.
For some objects that are also detected as \ha\ emitters (HAEs), dust attenuation can be independently estimated from the \ha/\pab\ ratio. 
However, such estimates are not available for the full galaxy sample. 
Therefore, in this paper we adopt the dust attenuation derived from the SED fitting in order to apply a uniform method to all galaxies. 
A detailed comparison between the SED-based attenuation and that inferred from the \ha/\pab\ ratio is presented in \cite{Jose2024a}.

\begin{table*}
\centering
 \caption{Input parameters for SED Fitting with CIGALE. We utilize a six-module, \textit{dustatt\_modified\_starburst module} \citep[][]{Calzetti2000s}, \textit{bc3} \citep{BruzualCharlot2003}, \textit{nebula}, \textit{dale2014}, \textit{Dust Emission}, \textit{redshifting}, produced by CIGALE.}
 \label{table:cigale}
  \begin{tabular}{lcc}\hline
  Parameters & \\\hline
  \quad IMF &  & \cite{Chabrier2003} \\
  \quad Age of the stellar population in the galaxy & & 10, 30, 50, 100, 300, 500, 1000, 3000 Myr \\
  \quad e-folding time of the stellar population model & & 100, 300, 500, 1000, 3000, 5000, 10000 Myr \\
  \quad Metallicity &  & 0.004 \\
  \quad Dust attenuation & & $0 < A_{\mathrm{v}} < 3$ \\
  \quad Nebular to continuum ratio &  & 1\\
  \quad $R_v$ &  & 4.05\\
  \quad Ionization parameter &  & -2.0, -2.1, -2.2, -2.3, -2.4, -2.5, -2.6, -2.7, -2.8\\
  \quad Gas metallicity &  & 0.004\\
  \quad Electron density &  & 100\\
  \quad Fraction of Lyman continuum photons escaping the galaxy &  & 0.0\\
  \quad Fraction of Lyman continuum photons absorbed by dust &  & 0.0\\
  \quad AGN fraction &  & 0.0\\\hline
  \end{tabular}
\end{table*}

\subsection{Star formation rate}
\label{subsec:SFR}
We derive SFR from \pab\ line flux measured from the NB image \citep{Bunker1995}.
The line flux can be derived as:
\begin{equation}
    F_{\mathrm{line}} = \frac{f_{\mathrm{F405N}}-f_{\mathrm{F410M}}}{1-\Delta_{\mathrm{F405N}}/\Delta_{\mathrm{F410M}}}\Delta_{\mathrm{F405N}},\label{eq:lineflux}
\end{equation}
in which $f_{\mathrm{F405N}}$ and $f_{\mathrm{F410M}}$ are the F405N and F410M flux densities, respectively, and $\Delta_{\mathrm{F405N}}$ and $\Delta_{\mathrm{F410M}}$ are the FWHMs of F405N and F410M.
We derive SFR from \pab\ line flux from Equation~\ref{eq:lineflux},
\begin{equation}
    \mathrm{SFR_{Pa\beta}}=8.48\times10^{-41}\left(\frac{L_{{\rm Pa}\beta,\,{\rm corr}}}{\mathrm{erg\,s^{-1}}}\right),\label{eq:SFR}
\end{equation}
where $L_{{\rm Pa}\beta,\,{\rm corr}}$ is the dust-corrected Pa$\beta$ luminosity.
This relation is obtained from the \citet{Kennicutt1998} H$\alpha$-based calibration, divided by a factor of 1.64 to convert from a Salpeter IMF to a \citet{Chabrier2003} IMF, and assuming an intrinsic H$\alpha$/Pa$\beta$ ratio of 17.6 under Case B recombination for $T_e = 10^4$ K and $n_e = 10^2\ {\rm cm^{-3}}$ \citep{Osterbrock2006}.
The H$\alpha$/Pa$\beta$ ratio is used only for the line conversion, while the dust attenuation correction is based on the attenuation derived from the SED fitting described in Section~3.2.

To correct for dust attenuation, we assume the attenuation curve of $A^{\prime}(\lambda)= k^{\prime}(\lambda) E(B-V)_{\mathrm{nebula}}$ \citep{Calzetti2000}.
The coefficient $k^{\prime}$ is defined shown as $k^{\prime}(\lambda) = 2.659(-1.857+1.040 / \lambda)+R_{V}^{\prime}$.
We also correct for galactic dust extinction using the NASA/IPAC Extragalactic Database extinction law calculator \footnote{http://irsa.ipac.caltech.edu/applications/DUST/}.
As noted in Section~\ref{subsec:sed}, we utilize the dust attenuation derived from the SED fit.

\section{Results}

\begin{figure}
\includegraphics[width=\columnwidth]{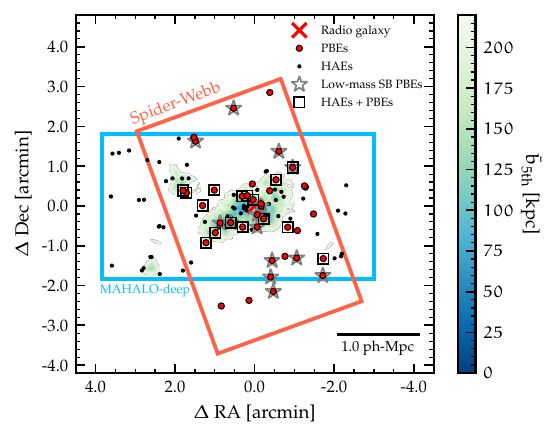}
\caption{The spatial distribution of the Spiderweb protocluster and Field of View (FoV) of the JWST/NIRCam (Spider-Webb project) and Subaru/MOIRCS (MAHALO-deep project; PI, T. Kodama) observation. The central red cross is the radio galaxy PKS1138-032, orange circles are PBEs, black dots are HAEs, squares are dual emitters (HAEs + PBEs), and magenta stars are low-mass PBEs ($M_{\star}<10^{9.0} \mathrm{M_{\odot}}$). Red and blue line show the FoV of Spider-Webb project and MAHALO-deep project. HAE samples are obtained from \citet{Daikuhara2024}. Colour bar represents the mean projected distance from 5th nearest HAEs. Low-mass galaxies in the Spiderweb, which exhibit relatively high SFR, are distributed entire region of the Spiderweb protocluster.}
\label{fig:dist}
\end{figure}
 
\begin{figure*}
  \begin{tabular}{cc}
    \begin{minipage}{0.50\textwidth}
      \begin{center}
        \includegraphics[width=\columnwidth]{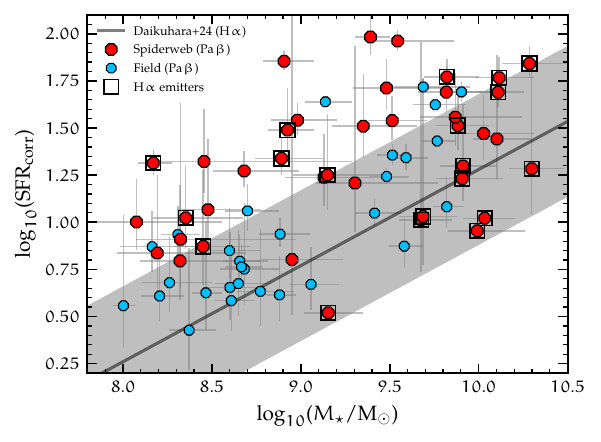}
      \end{center}
    \end{minipage}
    \begin{minipage}{0.50\textwidth}
      \begin{center}
        \includegraphics[width=\columnwidth]{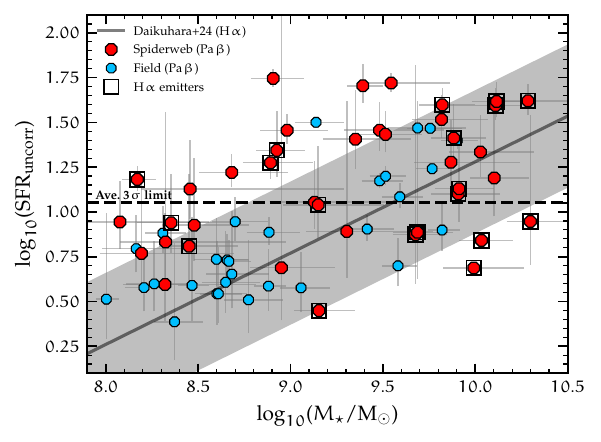}
      \end{center}
    \end{minipage}
  \end{tabular}
\caption{Dust-corrected SFR -- $M_{\star}$ (left) and dust-uncorrected SFR -- $M_{\star}$ (right) for the PBEs in the Spiderweb (red) and COSMOS Field \citep[blue; ][]{Duncan2025,Pirie2025}. The SFRs are derived from \pab\ line flux for PBEs and \ha\ line flux, respectively. The uncertainties on $M_{\star}$ are derived from the Bayesian estimates provided by CIGALE. The uncertainties on SFR come from photometric errors. Open squares denote HAEs as defined in \citet{Daikuhara2024}. The line shows MS of HAEs  \citep{Daikuhara2024}. Shaded areas indicates typical variation of $\pm$ 0.3 dex. We only plot Pa$\beta$ emitters with stellar masses $M_\star \ge 10^8\,M_\odot$, which corresponds to the stellar-mass detection limit of the Spiderweb Pa$\beta$ emitters. }
\label{fig:MS}
\end{figure*}

\begin{figure}
\centering
	\includegraphics[width=\columnwidth]{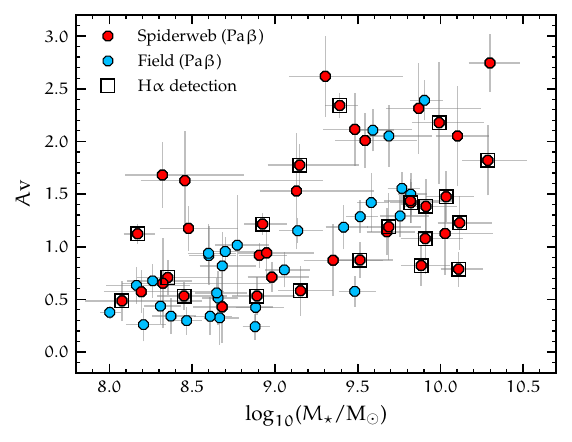}
\caption{The SED derived $A_V$ -- $M_{\mathrm{\star}}$ relation for PAEs and HAEs. $A_V$ and $M_{\mathrm{\star}}$ are derived from SED fitting. The uncertainties on $A_V$ and $M_{\star}$ are derived from the Bayesian estimates provided by CIGALE. The symbols correspond to those found in Figure~\ref{fig:MS}.}
\label{fig:hs17000_dustmass}
\end{figure}

\subsection{\pab\ emitters in Spiderweb protocluster}
We studied 40 PBEs excluding a radio galaxy (Spiderweb galaxy), and performed SED fitting to characterise their physical properties.
Cross-matching with a catalogue of HAEs yielded 19 PBEs with secure HAE counterparts within $0\farcs5$ \citep{Shimakawa2018pks,Daikuhara2024}.
By additionally requiring a $5\sigma$ detection in the \ha\ emission map to confirm PBEs, we identified 20 PBEs as \ha\ counterparts, implying a membership fraction of 0.55 (22/40).
Some galaxies are detected in Pa$\beta$ NB selection but not in H$\alpha$ NB selection. 
As discussed by \cite{Shimakawa2024a}, this discrepancy is primarily due to observational differences between the two surveys, including the incomplete spatial overlap, the slight offset in the NB redshift coverage, and the limited sensitivity to faint or blended sources in the seeing-limited ground-based H$\alpha$ NB data, rather than dust attenuation alone.

The \ha\ and \pab\ emission images (see Appendix~\ref{sec:B}) are constructed based on the F405N and F410M for \pab emission line and MOIRCS/NB2071 and MOIRCS/$K_s$ filters for \ha\ (+\nii) emission line using the following equation:
\begin{equation}
F_{\mathrm{line}}(x_i,y_i) = \frac{f_{\mathrm{NB}}(x_i,y_i)-f_{\mathrm{MB\,or\,BB}}(x_i,y_i)}{1-\Delta_{\mathrm{NB}}/\Delta_{\mathrm{MB\,or\,BB}}},\Delta_{\mathrm{NB}},\label{eq_line}
\end{equation}

Figure~\ref{fig:MS} shows the relation between SFR and $M_{\star}$ (the star-forming MS).
Our SFR detection limit corresponds to $11.25\,\mathrm{M_{\odot}\,yr^{-1}}$ in the dust-free case.
The Spiderweb PBEs span stellar masses of $M_{\star} = 10^{8.0}$–$10^{10.5}\,\mathrm{M_{\odot}}$, reaching $\sim1$ dex lower in mass than galaxies previously selected using MOIRCS-\ha\ NB imaging \citep{Shimakawa2018pks,Daikuhara2024}.
Notably, many massive galaxies ($>10^{10.5}\,\mathrm{M_{\odot}}$) known in the Spiderweb protocluster \citep{koyama2013,Shimakawa2018pks,Daikuhara2024} are not detected here, because their \pab\ EWs fall below our detection limit \citep{Shimakawa2024a}.
Compared to the MAHALO survey \citep{Daikuhara2024}, our smaller filter set increases the uncertainties in the SED-derived parameters.
For 71 \% of the PBEs, the stellar masses and dust-corrected SFRs are consistent within $\pm0.3$\, dex (Figure~\ref{fig:comp}).
Outliers may be affected by the presence of neighbouring sources or by observational uncertainties.
The consistency with \ha\ indicates that, although intrinsically faint, \pab\ can nevertheless serve as a reliable tracer of star-forming galaxies.

For comparison, we utilize field PBEs selected by the JWST Emission Line Survey (JELS; GO2321, P.I. P. Best) \citep{Duncan2025,Pirie2025} and the star-forming MS at $z\sim2$ derived from field \ha\ emitters \citep{Daikuhara2024}.
JELS obtained F466N and F470N narrow-band imaging in the COSMOS field.
Their dataset is composed of HST/F275W, F435W, F606W, F814W, F125W, F140W, F160W, JWST/F090W, F115W, F150W, F200W, F277W, F356W, F410M, F444W, F466N, and F470N.
To ensure consistency, both the protocluster and field samples are analysed using the same SED–fitting methodology applied to the same population of PBEs, thereby minimizing systematic uncertainties and enabling a fair assessment of environmental effects.
The survey depth corresponds to an SFR of $\sim3.7\ \mathrm{M_{\odot}\,yr^{-1}}$, and 35 PBE candidates were detected across 63 arcmin$^{2}$ \citep{Duncan2025,Pirie2025}.
PBEs in both the Spiderweb protocluster and the field are consistent with the MS derived from \ha\ emitters (Figure~\ref{fig:MS}), reinforcing that \pab\ is a reliable tracer of star-formation activity \citep[see also ][]{Jose2024a}.
Because the \pab\ line has a longer wavelength than \ha, it is less affected by dust attenuation, even though its intrinsic intensity is lower.

The SED derived $A_\mathrm{V}$ distribution adopted for dust corrections is shown in Figure~\ref{fig:hs17000_dustmass}.
While field and protocluster galaxies exhibit broadly similar distributions, we identify two low-mass galaxies ($M_{\star}\sim 10^{8.4}\,\mathrm{M_{\odot}}$) with high attenuation; one of these also shows an SFR $>11.25\, \mathrm{M_{\odot}\,yr^{-1}}$.
However, since this source was not detected in \ha\ emission, we cannot rule out contamination.

\subsection{Starbursting low-mass galaxies in Spiderweb protocluster}
Figure~\ref{fig:MS} highlights enhanced star-formation activity at the low-mass end ($M_{\star}<10^{9}\,\mathrm{M_{\odot}}$) within the Spiderweb protocluster with respect to field galaxies.
Despite the deeper NB imaging for the field, no low-mass PBEs with $\log(M_{\star}/\mathrm{M_\odot}) < 9.3$ are found in the field above the 3$\sigma$ SFR limit of the protocluster dataset ($11.25\,\mathrm{M_\odot\,yr^{-1}}$), revealing a clear excess of low-mass PBEs in the protocluster relative to the field.
This result suggests that the low-mass PBE population differs between the protocluster and field environments, consistent with the findings for the USS1558 protocluster at $z=2.53$ \citep{Hayashi2016,Daikuhara2024}.
Although a formal statistical comparison is not feasible if PBEs below the $3\sigma$ SFR limit are included, the observed excess of low-mass PBEs in the protocluster remains evident above the $3\sigma$ SFR limit.

The right panel of Figure~\ref{fig:MS} shows the SFR–$M_\star$ relation without dust correction.
In principle, the apparent enhancement of star-formation activity could arise from uncertainties in the SED fitting, particularly in estimates of stellar mass and dust attenuation.
However, even in this dust uncorrected diagram, the enhancement remains evident in the low-mass population, independent of SED-based dust attenuation.
These findings provide strong evidence that low-mass PBEs in the Spiderweb protocluster have elevated star formation compared to their field counterparts.

Around cosmic noon, previous studies have found weak or no environmental dependence of the star-forming MS at fixed stellar mass \citep[e.g.,][]{koyama2013, Darvish2016, Jose2023}.
In contrast, we find that only low-mass PBEs in the Spiderweb protocluster lie above the field MS, whereas more massive galaxies are fully consistent with the field relation.
This indicates that the environmental effect on the MS has a strong mass dependence and suggests that previous studies, which mainly focused on massive galaxies, may have missed a population of environmentally enhanced low-mass star-forming galaxies.

It should also be noted that previous \ha\ NB surveys have missed dusty HAEs.
Nevertheless, while our \pab\ observations revealed many galaxies that were undetected in \ha\ emitter survey \citep{Shimakawa2018pks}, our overall trends remain consistent with previous \ha\ emitter survey results and do not contradict earlier findings \citep{koyama2013,Shimakawa2018pks,Daikuhara2024}.

\section{Discussion}

\begin{table*}
\centering
\renewcommand{\arraystretch}{1.5} 
\setlength{\tabcolsep}{10pt} 
\caption{Summary of environmental effects on galaxies in two protoclusters (Spiderweb and USS1558) classified by stellar mass. The definitions of the stellar-mass bins are based on \citet{Daikuhara2024}. "Enhanced" indicates the presence of environmental influences on star formation activity.}
\label{tab:summary}
\begin{tabular}{c|ccc}
\hline
\textbf{Environment} & \textbf{High-mass} & \textbf{Intermediate-mass} & \textbf{Low-mass} \\
 & $M_{\star}/\mathrm{M_{\odot}}>10^{10}$&  $10^{10}>M_{\star}/\mathrm{M_{\odot}}>10^{9}$ & $10^{9}>M_{\star}/\mathrm{M_{\odot}}>10^{8}$ \\
\hline
Spiderweb & No & No & \textbf{Enhanced} \\
\hline
USS1558$-$003 & No & \textbf{Enhanced} (only in the core) & \textbf{Enhanced} \\
\hline
\end{tabular}
\end{table*}
\subsection{Mass and environmental dependence of enhanced star formation}
Intermediate-to-high-mass PBEs ($M_\star/M_\odot > 10^{9}$) do not show significant enhancement in SFR in the Spiderweb protocluster, as confirmed by a Kolmogorov–Smirnov test ($p>0.05$).
This is consistent with \ha\ emitter survey in the Spiderweb \citep{Shimakawa2018pks,Daikuhara2024}.
However, this result is different from \ha\ emitters in the USS1558–003 protocluster, where enhanced star formation is found in intermediate-mass galaxies, although only in the dense core region.
The contrast between the two systems suggests that enhanced star formation depends on the dynamical stage of the protocluster (Table~\ref{tab:summary}).
The Spiderweb protocluster represents a mature evolutionary stage among protoclusters at cosmic noon \citep{Shimakawa2018pks}.
The detection of the thermal Sunyaev–Zeldovich (tSZ) effect \citep{DiMascolo2023} indicates the presence of a hot intracluster medium, suggesting that the system has transitioned to the hot-mode of gas accretion \citep{Dekel2009}.  
In spite of this, star-forming galaxies still follow the same MS as those in the field, implying that their star formation is governed by common, self-regulated physical processes independent of environment \citep[see also ][]{Shimakawa2018pks,Jose2023,Daikuhara2024}.
The higher fraction of quiescent galaxies, however, suggests that the quenching process is already acting on galaxies in the Spiderweb protocluster \citep{Naufal2024}.

In contrast, we find that low-mass galaxies in the Spiderweb protocluster exhibit enhanced star formation, as traced by Pa$\beta$, compared with their field counterparts.
The spatial distribution of starbursting low-mass galaxies shows no concentration around the central radio galaxy (Figure~\ref{fig:dist}). 
The scattered locations of starbursting galaxies could also reflect mergers/interactions and/or recently accreted galaxies influenced by early environmental effects \citep[i.e., pre-processing; e.g.,][]{Zabludoff1998,Fujita2004}.
Cold accretion may still operate in the outskirts where the intracluster medium is not yet fully virialized.
Ram-pressure compression may also contribute  \citep{Bekki2003,Dannerbauer2017,Vulcani2018,George2025}. 
However, because the star-bursting low-mass galaxies do not cluster toward the densest region, a ram-pressure-driven mechanism is unlikely to be the primary driver in our sample.

Hydrodynamic simulations, TNG100 and TNG300, predict that such recent mergers/interactions can temporarily enhance star formation \citep[][Hagiwara et al. in preparation]{Andrews2024}, which is qualitatively consistent with our findings.
Although the internal structures of these low-mass galaxies are unresolved, previous studies report a high merger fraction in the Spiderweb protocluster \citep{Naufal2023} and the presence of compact star-forming systems \citep{Daikuhara2024}.
These characteristics may indicate that some galaxies are undergoing centrally concentrated starbursts triggered by interactions/mergers and/or violent disc instability.

Despite these differences in the physical conditions of the ICM between the two protoclusters, low-mass galaxies in both protoclusters show comparable SFRs.
This contrast implies that the mechanisms driving star-formation enhancement differ between low- and intermediate-mass galaxies.
For intermediate-mass galaxies, enhanced star formation may depend on the dynamical state of the protocluster, which likely regulates the efficiency of large-scale cold-gas supply.
In contrast, low-mass galaxies may be less tightly coupled to the global ICM conditions and can experience elevated star formation across a wide range of protocluster environments through more local and stochastic processes, such as pre-processing in infalling groups, galaxy mergers/interactions, and efficient cold-gas accretion from surrounding filaments.
Overall, this mass-dependent behavior points to multiple, distinct pathways through which the environment facilitates star formation during protocluster assembly.

\subsection{Star formation activity of low-mass galaxies in protoclusters at cosmic noon}
We do not fully capture the MS at low mass because our Pa$\beta$ observations do not reach sufficient depth to sample star-forming galaxies on the sequence.
Thus, it remains uncertain whether the star formation activity of low-mass protocluster galaxies is elevated across the whole MS, or whether only a subset of galaxies are undergoing short-term starburst events ($<$100 Myr).
HST observations \citep{Santini2017} revealed that the scatter of the MS increases toward lower masses, although the presence and strength of this mass dependence remain debated. 
For example, \citet{Kurczynski2016} found no significant SFR enhancement at the low-mass end over $0.5<z<3$ and $10^7$ -- $10^{11}\,M_\odot$. 
They also argued that the scatter could become larger for low-mass galaxies on a short timescale ($<$100 Myr). 

Pa$\beta$ traces star formation on a short timescale \citep[$<$10 Myr; e.g., ][]{Kennicutt2012,Madau2014,daCunha2016}.
If environmental processes in protoclusters trigger such short-timescale starbursts, this could naturally lead to an apparent excess of starbursting low-mass galaxies in our sample compared to the field.
Theoretical works support this view, showing that fluctuations in the SFR–$M_\star$ relation arise from both long- and short-term variations, with short-timescale changes being especially important for low-mass galaxies \citep{Matthee2019}.
If these environmentally driven bursts are common, they could contribute to the increased scatter of the star-forming MS observed in dense environments.
However, the relative importance and characteristic timescales of these processes remain uncertain.
Further studies that probe a wider range of SFRs and utilize several star formation tracers will be necessary to clarify how short-term variability and environment together shape the star-forming MS.

In the local Universe, dense environments is well established as a key factor to suppress star formation activity \citep[][]{Dressler1980,Butcher1984,Bamford2009,Cappellari2011,vanderWel2008,Bait2017}.  
On the other hand, dense environments at cosmic noon can facilitate more star formation. Such 
an enhancement depends on galaxy mass and the dynamical state of protoclusters and may be driven by efficient cold gas supply and/or frequent galaxy–galaxy interactions.
For massive galaxies, star formation remains largely self-regulated by internal processes with little environmental dependence \citep{koyama2013,Hayashi2016}, whereas low-mass galaxies are more sensitive to external influences.  
This mass-dependent response is consistent with the trends observed also in the local universe \citep{Peng2010}, suggesting that the susceptibility of low-mass galaxies to their environment is a persistent feature of galaxy evolution across cosmic time.

\section{Conclusions}

We investigated the star formation activity of Pa$\beta$ emitters in the Spiderweb protocluster at $z = 2.16$, using a JWST NB-selected sample  \citep{Shimakawa2024a} that is sensitive to recent star formation on short timescales.
SFRs were estimated from the Pa$\beta$ emission-line luminosities under the assumption of the case~B recombination.  
Other physical properties, such as stellar mass and dust attenuation, were derived through SED fitting using \textit{JWST}/NIRCam and \textit{HST} imaging data.  
For field comparison, we employed PBEs from the \textit{JWST} Emission Line Survey in the COSMOS field  \citep{Duncan2025,Pirie2025}.  
Both protocluster and field samples were analysed using the same SED-fitting procedure to ensure a consistent and unbiased comparison.  
We summarize our main results and discussion below.

\begin{itemize}
    \item \textbf{Enhanced star formation in low-mass PBEs:}  
    Our main finding is that low-mass PBEs ($M_\star < 10^9\,M_\odot$) in the Spiderweb protocluster exhibit a clear and significant enhancement in star formation compared to their field counterparts.  
    This excess remains even without applying dust-attenuation corrections, confirming that it is not an artifact of SED-fitting uncertainties. 
    In contrast, intermediate- and high-mass PBEs show no significant deviation from the field, revealing a strong mass dependence in the environmental effects on star formation.

    \item \textbf{Comparison with USS1558--003 protocluster:}  
    A comparison with the USS1558--003 protocluster at $z = 2.53$ \citep{Hayashi2016,Daikuhara2024} reveals similar behavior in low-mass galaxies but different trends for intermediate-mass galaxies.  
    These contrasting patterns reflect the distinct dynamical stages of the two protoclusters: USS1558--003 is in an early assembly phase where cold gas accretion remains efficient, whereas the Spiderweb protocluster is a more evolved structure with a hot ICM as evidenced by the thermal Sunyaev--Zeldovich effect \citep{DiMascolo2023}.  
    However, low-mass galaxies in both protoclusters show common trends in spatial distribution and SFRs. 
    This may suggest that similar environmental mechanisms are at work in regulating the evolution of low-mass galaxies.

    \item \textbf{Starbursting mechanisms of low-mass galaxies:}  
    Starbursting low-mass galaxies are not clustered around the central radio galaxy but are distributed across the protocluster region.  
    This extended spatial distribution may indicate that environmental mechanisms such as galaxy mergers/interactions and/or cold gas accretion facilitate star formation.

    \item \textbf{Mass and dynamical stage dependence of enhanced star formation activity:}  
    Using a uniform SED-fitting approach for both protocluster and field PBEs minimizes systematic uncertainties and ensures the robustness of these conclusions.
    Our results indicate that the role of environment at cosmic noon depends jointly on galaxy mass and the dynamical stage of the protocluster.  
    Intermediate-mass galaxies are more sensitive to large-scale gas inflows in systems where cold accretion is still active, while low-mass galaxies are more strongly affected by interactions/mergers and/or cold gas accretion.
\end{itemize}

\section*{Acknowledgements}
We thank the anonymous referee for the useful comments.
This work is based on observations made with the NASA/ ESA/CSA James Webb Space Telescope. 
The data were obtained from the Barbara A. Mikulski Archive for Space Telescopes at the Space Telescope Science Institute at doi:10.17909/vx25-q902, which is operated by the Association of Universities for Research in Astronomy, Inc., under NASA contract NAS 5-03127 for JWST. 
The observation is associated with programme No. 1572 in cycle 1.
This research is based in part on observations made with the NASA/ESA Hubble Space Telescope obtained from the Space Telescope Science Institute, which is operated by the Association of Universities for Research in Astronomy, Inc., under NASA contract NAS 5-26555.
HST observations are associated with programme ID 10327 and ID 17117.
This research is based in part on data collected at the Subaru Telescope, which is operated by the National Astronomical Observatory of Japan. 
We are honored and grateful for the opportunity to observe the Universe from Maunakea, which is cultural, historical, and natural significance in Hawaii.
This research made use of Astropy \citep{astropy:2013,astropy:2018,astropy:2022}, NumPy \citep{harris2020array}, Matplotlib \citep{Hunter2007}, TOPCAT \citep{Taylor2011}, SExtractor \citep{Bertin1996}, Photutils \citep{Photutils}, SWarp \citep{Bertin2002,Bertin2010}, and SCAMP \citep{Bertin2006}.

This work was supported by JSPS KAKENHI Grant Numbers 25K23411 (Research Activity Start-up by K. Daikuhara), JST SPRING, Japan Grant Number JPMJSP2114, JSPS Core-to-Core Program (JPJSCCA20210003), and Graduate Program on Physics for the Universe (GP-PU) Tohoku University.
TK acknowledges financial support from JSPS KAKENHI Grant Numbers 24H00002 (Specially Promoted Research by T. Kodama et al.) and 22K21349 (International Leading Research by S. Miyazaki et al.).
YK acknowledge financial support from JSPS KAKENHI Grant Numbers 24H00002 (Specially Promoted Research by T. Kodama et al.), 22K21349 (International Leading Research by S. Miyazaki et al.), 23H01219 (Kiban-B by Y. Koyama et al.), 26H02070 (Kiban-A by Y. Koyama et al.), and JSPS Core-to-Core Program (JPJSCCA20210003; M. Yoshida et al.).
JMPM acknowledges that this project has received funding from the European Union’s Horizon research and innovation programme under the Marie Skłodowska-Curie grant agreement No 101106626.
HD and JMPM acknowledge support from the Agencia Estatal de 820 Investigación del Ministerio de Ciencia, Innovación y Universidades (MCIU/AEI) under grant (Construcción de cúmulos de galaxias en formación a través de la formación estelar oscurecida por el polvo) and the European Regional Development Fund (ERDF) with reference (PID2022-143243NB-I00/DOI:10.13039/501100011033).
EI gratefully acknowledge financial support from ANID - MILENIO - NCN2024\_112 and ANID FONDECYT Regular 1221846.

\section*{Data Availability}
All JWST raw frames are available from the Barbara A. Mikulski Archive for Space Telescopes at doi:10.17909/vx25-q902 associated with programme No. 1572 in cycle 1.
HST raw frame are associated with programme ID 10327 and ID 17117.
The corresponding reduced data products and a photometric catalogue, including the NB-selected star-forming galaxies in this field, can be obtained from the authors upon reasonable request.



\bibliographystyle{mnras}
\bibliography{Main} 

@INPROCEEDINGS{Kodama2013,
       author = {{Kodama}, Tadayuki and {Hayashi}, Masao and {Koyama}, Yusei and {Tadaki}, Ken-ichi and {Tanaka}, Ichi and {Shimakawa}, Rhythm},
        title = "{Mahalo-Subaru: Mapping Star Formation at the Peak Epoch of Massive Galaxy Formation}",
    booktitle = {The Intriguing Life of Massive Galaxies},
         year = 2013,
       editor = {{Thomas}, Daniel and {Pasquali}, Anna and {Ferreras}, Ignacio},
       series = {Proceedings of the International Astronomical Union},
       volume = {295},
        month = jul,
        pages = {74-77},
          doi = {10.1017/S1743921313004353},
       adsurl = {https://ui.adsabs.harvard.edu/abs/2013IAUS..295...74K}
}

@ARTICLE{Santini2017,
       author = {{Santini}, Paola and {Fontana}, Adriano and {Castellano}, Marco and {Di Criscienzo}, Marcella and {Merlin}, Emiliano and {Amorin}, Ricardo and {Cullen}, Fergus and {Daddi}, Emanuele and {Dickinson}, Mark and {Dunlop}, James S. and {Grazian}, Andrea and {Lamastra}, Alessandra and {McLure}, Ross J. and {Micha{\l}owski}, Micha{\l}. J. and {Pentericci}, Laura and {Shu}, Xinwen},
        title = "{The Star Formation Main Sequence in the Hubble Space Telescope Frontier Fields}",
      journal = {\apj},
         year = 2017,
        month = sep,
       volume = {847},
       number = {1},
          eid = {76},
        pages = {76},
          doi = {10.3847/1538-4357/aa8874},
archivePrefix = {arXiv},
       eprint = {1706.07059},
 primaryClass = {astro-ph.GA},
       adsurl = {https://ui.adsabs.harvard.edu/abs/2017ApJ...847...76S}
}

@ARTICLE{Shimakawa2014,
       author = {{Shimakawa}, R. and {Kodama}, T. and {Tadaki}, K. -I. and {Tanaka}, I. and {Hayashi}, M. and {Koyama}, Y.},
        title = "{Identification of the progenitors of rich clusters and member galaxies in rapid formation at z > 2.}",
      journal = {\mnras},
         year = 2014,
        month = jun,
       volume = {441},
        pages = {L1-L5},
          doi = {10.1093/mnrasl/slu029},
archivePrefix = {arXiv},
       eprint = {1402.3568},
 primaryClass = {astro-ph.GA},
       adsurl = {https://ui.adsabs.harvard.edu/abs/2014MNRAS.441L...1S}
}

@ARTICLE{Hayashi2016,
       author = {{Hayashi}, Masao and {Kodama}, Tadayuki and {Tanaka}, Ichi and {Shimakawa}, Rhythm and {Koyama}, Yusei and {Tadaki}, Ken-ichi and {Suzuki}, Tomoko L. and {Yamamoto}, Moegi},
        title = "{Enhanced Star Formation of Less Massive Galaxies in a Protocluster at z = 2.5}",
      journal = {\apjl},
         year = 2016,
        month = aug,
       volume = {826},
       number = {2},
          eid = {L28},
        pages = {L28},
          doi = {10.3847/2041-8205/826/2/L28},
archivePrefix = {arXiv},
       eprint = {1607.04040},
 primaryClass = {astro-ph.GA},
       adsurl = {https://ui.adsabs.harvard.edu/abs/2016ApJ...826L..28H}
}

@ARTICLE{Shimakawa2018pks,
       author = {{Shimakawa}, Rhythm and {Koyama}, Yusei and {R{\"o}ttgering}, Huub J.~A. and {Kodama}, Tadayuki and {Hayashi}, Masao and {Hatch}, Nina A. and {Dannerbauer}, Helmut and {Tanaka}, Ichi and {Tadaki}, Ken-ichi and {Suzuki}, Tomoko L. and {Fukagawa}, Nao and {Cai}, Zheng and {Kurk}, Jaron D.},
        title = "{MAHALO Deep Cluster Survey II. Characterizing massive forming galaxies in the Spiderweb protocluster at z = 2.2}",
      journal = {\mnras},
         year = 2018,
        month = dec,
       volume = {481},
       number = {4},
        pages = {5630-5650},
          doi = {10.1093/mnras/sty2618},
archivePrefix = {arXiv},
       eprint = {1809.08755},
 primaryClass = {astro-ph.GA},
       adsurl = {https://ui.adsabs.harvard.edu/abs/2018MNRAS.481.5630S}
}

@ARTICLE{Calzetti2000s,
       author = {{Calzetti}, Daniela and {Armus}, Lee and {Bohlin}, Ralph C. and {Kinney}, Anne L. and {Koornneef}, Jan and {Storchi-Bergmann}, Thaisa},
        title = "{The Dust Content and Opacity of Actively Star-forming Galaxies}",
      journal = {\apj},
         year = 2000,
        month = apr,
       volume = {533},
       number = {2},
        pages = {682-695},
          doi = {10.1086/308692},
archivePrefix = {arXiv},
       eprint = {astro-ph/9911459},
 primaryClass = {astro-ph},
       adsurl = {https://ui.adsabs.harvard.edu/abs/2000ApJ...533..682C}
}

@ARTICLE{Chabrier2003,
       author = {{Chabrier}, Gilles},
        title = "{Galactic Stellar and Substellar Initial Mass Function}",
      journal = {\pasp},
         year = 2003,
        month = jul,
       volume = {115},
       number = {809},
        pages = {763-795},
          doi = {10.1086/376392},
archivePrefix = {arXiv},
       eprint = {astro-ph/0304382},
 primaryClass = {astro-ph},
       adsurl = {https://ui.adsabs.harvard.edu/abs/2003PASP..115..763C}
}

@ARTICLE{Bertin1996,
       author = {{Bertin}, E. and {Arnouts}, S.},
        title = "{SExtractor: Software for source extraction.}",
      journal = {\aaps},
         year = 1996,
        month = jun,
       volume = {117},
        pages = {393-404},
          doi = {10.1051/aas:1996164},
       adsurl = {https://ui.adsabs.harvard.edu/abs/1996A&AS..117..393B}
}

@ARTICLE{BruzualCharlot2003,
       author = {{Bruzual}, G. and {Charlot}, S.},
        title = "{Stellar population synthesis at the resolution of 2003}",
      journal = {\mnras},
         year = 2003,
        month = oct,
       volume = {344},
       number = {4},
        pages = {1000-1028},
          doi = {10.1046/j.1365-8711.2003.06897.x},
archivePrefix = {arXiv},
       eprint = {astro-ph/0309134},
 primaryClass = {astro-ph},
       adsurl = {https://ui.adsabs.harvard.edu/abs/2003MNRAS.344.1000B}
}

@ARTICLE{koyama2014,
       author = {{Koyama}, Yusei and {Kodama}, Tadayuki and {Tadaki}, Ken-ichi and {Hayashi}, Masao and {Tanaka}, Ichi and {Shimakawa}, Rhythm},
        title = "{The Environmental Impacts on the Star Formation Main Sequence: An H{\ensuremath{\alpha}} Study of the Newly Discovered Rich Cluster at z = 1.52}",
      journal = {\apj},
         year = 2014,
        month = jul,
       volume = {789},
       number = {1},
          eid = {18},
        pages = {18},
          doi = {10.1088/0004-637X/789/1/18},
archivePrefix = {arXiv},
       eprint = {1405.4165},
 primaryClass = {astro-ph.GA},
       adsurl = {https://ui.adsabs.harvard.edu/abs/2014ApJ...789...18K}
}

@ARTICLE{Overzier2016,
       author = {{Overzier}, Roderik A.},
        title = "{The realm of the galaxy protoclusters. A review}",
      journal = {\aapr},
         year = 2016,
        month = nov,
       volume = {24},
       number = {1},
          eid = {14},
        pages = {14},
          doi = {10.1007/s00159-016-0100-3},
archivePrefix = {arXiv},
       eprint = {1610.05201},
 primaryClass = {astro-ph.GA},
       adsurl = {https://ui.adsabs.harvard.edu/abs/2016A&ARv..24...14O}
}

@ARTICLE{koyama2013,
       author = {{Koyama}, Yusei and {Kodama}, Tadayuki and {Tadaki}, Ken-ichi and {Hayashi}, Masao and {Tanaka}, Masayuki and {Smail}, Ian and {Tanaka}, Ichi and {Kurk}, Jaron},
        title = "{Massive starburst galaxies in a z = 2.16 proto-cluster unveiled by panoramic H{\ensuremath{\alpha}} mapping}",
      journal = {\mnras},
         year = 2013,
        month = jan,
       volume = {428},
       number = {2},
        pages = {1551-1564},
          doi = {10.1093/mnras/sts133},
archivePrefix = {arXiv},
       eprint = {1210.0972},
 primaryClass = {astro-ph.CO},
       adsurl = {https://ui.adsabs.harvard.edu/abs/2013MNRAS.428.1551K}
}

@ARTICLE{Miley2006,
       author = {{Miley}, George K. and {Overzier}, Roderik A. and {Zirm}, Andrew W. and {Ford}, Holland C. and {Kurk}, Jaron and {Pentericci}, Laura and {Blakeslee}, John P. and {Franx}, Marijn and {Illingworth}, Garth D. and {Postman}, Marc and {Rosati}, Piero and {R{\"o}ttgering}, Huub J.~A. and {Venemans}, Bram P. and {Helder}, Eveline},
        title = "{The Spiderweb Galaxy: A Forming Massive Cluster Galaxy at z \raisebox{-0.5ex}\textasciitilde 2}",
      journal = {\apjl},
         year = 2006,
        month = oct,
       volume = {650},
       number = {1},
        pages = {L29-L32},
          doi = {10.1086/508534},
archivePrefix = {arXiv},
       eprint = {astro-ph/0610909},
 primaryClass = {astro-ph},
       adsurl = {https://ui.adsabs.harvard.edu/abs/2006ApJ...650L..29M}
}

@ARTICLE{Seymour2007,
       author = {{Seymour}, Nick and {Stern}, Daniel and {De Breuck}, Carlos and {Vernet}, Joel and {Rettura}, Alessandro and {Dickinson}, Mark and {Dey}, Arjun and {Eisenhardt}, Peter and {Fosbury}, Robert and {Lacy}, Mark and {McCarthy}, Pat and {Miley}, George and {Rocca-Volmerange}, Brigitte and {R{\"o}ttgering}, Huub and {Stanford}, S. Adam and {Teplitz}, Harry and {van Breugel}, Wil and {Zirm}, Andrew},
        title = "{The Massive Hosts of Radio Galaxies across Cosmic Time}",
      journal = {\apjs},
         year = 2007,
        month = aug,
       volume = {171},
       number = {2},
        pages = {353-375},
          doi = {10.1086/517887},
archivePrefix = {arXiv},
       eprint = {astro-ph/0703224},
 primaryClass = {astro-ph},
       adsurl = {https://ui.adsabs.harvard.edu/abs/2007ApJS..171..353S}
}

@ARTICLE{vanderBurg2020,
       author = {{van der Burg}, Remco F.~J. and {Rudnick}, Gregory and {Balogh}, Michael L. and {Muzzin}, Adam and {Lidman}, Chris and {Old}, Lyndsay J. and {Shipley}, Heath and {Gilbank}, David and {McGee}, Sean and {Biviano}, Andrea and {Cerulo}, Pierluigi and {Chan}, Jeffrey C.~C. and {Cooper}, Michael and {De Lucia}, Gabriella and {Demarco}, Ricardo and {Forrest}, Ben and {Gwyn}, Stephen and {Jablonka}, Pascale and {Kukstas}, Egidijus and {Marchesini}, Danilo and {Nantais}, Julie and {Noble}, Allison and {Pintos-Castro}, Irene and {Poggianti}, Bianca and {Reeves}, Andrew M.~M. and {Stefanon}, Mauro and {Vulcani}, Benedetta and {Webb}, Kristi and {Wilson}, Gillian and {Yee}, Howard and {Zaritsky}, Dennis},
        title = "{The GOGREEN Survey: A deep stellar mass function of cluster galaxies at 1.0 < z < 1.4 and the complex nature of satellite quenching}",
      journal = {\aap},
         year = 2020,
        month = jun,
       volume = {638},
          eid = {A112},
        pages = {A112},
          doi = {10.1051/0004-6361/202037754},
archivePrefix = {arXiv},
       eprint = {2004.10757},
 primaryClass = {astro-ph.GA},
       adsurl = {https://ui.adsabs.harvard.edu/abs/2020A&A...638A.112V}
}

@ARTICLE{Madau2014,
       author = {{Madau}, Piero and {Dickinson}, Mark},
        title = "{Cosmic Star-Formation History}",
      journal = {\araa},
         year = 2014,
        month = aug,
       volume = {52},
        pages = {415-486},
          doi = {10.1146/annurev-astro-081811-125615},
archivePrefix = {arXiv},
       eprint = {1403.0007},
 primaryClass = {astro-ph.CO},
       adsurl = {https://ui.adsabs.harvard.edu/abs/2014ARA&A..52..415M}
}

@ARTICLE{Shivaei2015,
       author = {{Shivaei}, Irene and {Reddy}, Naveen A. and {Shapley}, Alice E. and {Kriek}, Mariska and {Siana}, Brian and {Mobasher}, Bahram and {Coil}, Alison L. and {Freeman}, William R. and {Sanders}, Ryan and {Price}, Sedona H. and {de Groot}, Laura and {Azadi}, Mojegan},
        title = "{The MOSDEF Survey: Dissecting the Star Formation Rate versus Stellar Mass Relation Using H{\ensuremath{\alpha}} and H{\ensuremath{\beta}} Emission Lines at z {\ensuremath{\sim}} 2}",
      journal = {\apj},
         year = 2015,
        month = dec,
       volume = {815},
       number = {2},
          eid = {98},
        pages = {98},
          doi = {10.1088/0004-637X/815/2/98},
archivePrefix = {arXiv},
       eprint = {1507.03017},
 primaryClass = {astro-ph.GA},
       adsurl = {https://ui.adsabs.harvard.edu/abs/2015ApJ...815...98S}
}

@ARTICLE{Schreiber2015,
       author = {{Schreiber}, C. and {Pannella}, M. and {Elbaz}, D. and {B{\'e}thermin}, M. and {Inami}, H. and {Dickinson}, M. and {Magnelli}, B. and {Wang}, T. and {Aussel}, H. and {Daddi}, E. and {Juneau}, S. and {Shu}, X. and {Sargent}, M.~T. and {Buat}, V. and {Faber}, S.~M. and {Ferguson}, H.~C. and {Giavalisco}, M. and {Koekemoer}, A.~M. and {Magdis}, G. and {Morrison}, G.~E. and {Papovich}, C. and {Santini}, P. and {Scott}, D.},
        title = "{The Herschel view of the dominant mode of galaxy growth from z = 4 to the present day}",
      journal = {\aap},
         year = 2015,
        month = mar,
       volume = {575},
          eid = {A74},
        pages = {A74},
          doi = {10.1051/0004-6361/201425017},
archivePrefix = {arXiv},
       eprint = {1409.5433},
 primaryClass = {astro-ph.GA},
       adsurl = {https://ui.adsabs.harvard.edu/abs/2015A&A...575A..74S}
}

@ARTICLE{Whitaker2012,
       author = {{Whitaker}, Katherine E. and {van Dokkum}, Pieter G. and {Brammer}, Gabriel and {Franx}, Marijn},
        title = "{The Star Formation Mass Sequence Out to z = 2.5}",
      journal = {\apjl},
         year = 2012,
        month = aug,
       volume = {754},
       number = {2},
          eid = {L29},
        pages = {L29},
          doi = {10.1088/2041-8205/754/2/L29},
archivePrefix = {arXiv},
       eprint = {1205.0547},
 primaryClass = {astro-ph.CO},
       adsurl = {https://ui.adsabs.harvard.edu/abs/2012ApJ...754L..29W}
}

@article{Keres2005,
    author = {Kereš, Dušan and Katz, Neal and Weinberg, David H. and Davé, Romeel},
    title = "{How do galaxies get their gas?}",
    journal = {Monthly Notices of the Royal Astronomical Society},
    volume = {363},
    number = {1},
    pages = {2-28},
    year = {2005},
    month = {10},
    issn = {0035-8711},
    doi = {10.1111/j.1365-2966.2005.09451.x},
    url = {https://doi.org/10.1111/j.1365-2966.2005.09451.x},
    eprint = {https://academic.oup.com/mnras/article-pdf/363/1/2/4126209/363-1-2.pdf},
}

@ARTICLE{Noeske2007,
       author = {{Noeske}, K.~G. and {Weiner}, B.~J. and {Faber}, S.~M. and {Papovich}, C. and {Koo}, D.~C. and {Somerville}, R.~S. and {Bundy}, K. and {Conselice}, C.~J. and {Newman}, J.~A. and {Schiminovich}, D. and {Le Floc'h}, E. and {Coil}, A.~L. and {Rieke}, G.~H. and {Lotz}, J.~M. and {Primack}, J.~R. and {Barmby}, P. and {Cooper}, M.~C. and {Davis}, M. and {Ellis}, R.~S. and {Fazio}, G.~G. and {Guhathakurta}, P. and {Huang}, J. and {Kassin}, S.~A. and {Martin}, D.~C. and {Phillips}, A.~C. and {Rich}, R.~M. and {Small}, T.~A. and {Willmer}, C.~N.~A. and {Wilson}, G.},
        title = "{Star Formation in AEGIS Field Galaxies since z=1.1: The Dominance of Gradually Declining Star Formation, and the Main Sequence of Star-forming Galaxies}",
      journal = {\apjl},
         year = 2007,
        month = may,
       volume = {660},
       number = {1},
        pages = {L43-L46},
          doi = {10.1086/517926},
archivePrefix = {arXiv},
       eprint = {astro-ph/0701924},
 primaryClass = {astro-ph},
       adsurl = {https://ui.adsabs.harvard.edu/abs/2007ApJ...660L..43N}
}

@ARTICLE{Brinchmann2004,
       author = {{Brinchmann}, J. and {Charlot}, S. and {White}, S.~D.~M. and {Tremonti}, C. and {Kauffmann}, G. and {Heckman}, T. and {Brinkmann}, J.},
        title = "{The physical properties of star-forming galaxies in the low-redshift Universe}",
      journal = {\mnras},
         year = 2004,
        month = jul,
       volume = {351},
       number = {4},
        pages = {1151-1179},
          doi = {10.1111/j.1365-2966.2004.07881.x},
archivePrefix = {arXiv},
       eprint = {astro-ph/0311060},
 primaryClass = {astro-ph},
       adsurl = {https://ui.adsabs.harvard.edu/abs/2004MNRAS.351.1151B}
}

@ARTICLE{Daddi2007,
       author = {{Daddi}, E. and {Dickinson}, M. and {Morrison}, G. and {Chary}, R. and {Cimatti}, A. and {Elbaz}, D. and {Frayer}, D. and {Renzini}, A. and {Pope}, A. and {Alexander}, D.~M. and {Bauer}, F.~E. and {Giavalisco}, M. and {Huynh}, M. and {Kurk}, J. and {Mignoli}, M.},
        title = "{Multiwavelength Study of Massive Galaxies at z\raisebox{-0.5ex}\textasciitilde2. I. Star Formation and Galaxy Growth}",
      journal = {\apj},
         year = 2007,
        month = nov,
       volume = {670},
       number = {1},
        pages = {156-172},
          doi = {10.1086/521818},
archivePrefix = {arXiv},
       eprint = {0705.2831},
 primaryClass = {astro-ph},
       adsurl = {https://ui.adsabs.harvard.edu/abs/2007ApJ...670..156D}
}

@ARTICLE{Elbaz2007,
       author = {{Elbaz}, D. and {Daddi}, E. and {Le Borgne}, D. and {Dickinson}, M. and {Alexander}, D.~M. and {Chary}, R. -R. and {Starck}, J. -L. and {Brandt}, W.~N. and {Kitzbichler}, M. and {MacDonald}, E. and {Nonino}, M. and {Popesso}, P. and {Stern}, D. and {Vanzella}, E.},
        title = "{The reversal of the star formation-density relation in the distant universe}",
      journal = {\aap},
         year = 2007,
        month = jun,
       volume = {468},
       number = {1},
        pages = {33-48},
          doi = {10.1051/0004-6361:20077525},
archivePrefix = {arXiv},
       eprint = {astro-ph/0703653},
 primaryClass = {astro-ph},
       adsurl = {https://ui.adsabs.harvard.edu/abs/2007A&A...468...33E}
}

@ARTICLE{Salim2007,
       author = {{Salim}, Samir and {Rich}, R. Michael and {Charlot}, St{\'e}phane and {Brinchmann}, Jarle and {Johnson}, Benjamin D. and {Schiminovich}, David and {Seibert}, Mark and {Mallery}, Ryan and {Heckman}, Timothy M. and {Forster}, Karl and {Friedman}, Peter G. and {Martin}, D. Christopher and {Morrissey}, Patrick and {Neff}, Susan G. and {Small}, Todd and {Wyder}, Ted K. and {Bianchi}, Luciana and {Donas}, Jos{\'e} and {Lee}, Young-Wook and {Madore}, Barry F. and {Milliard}, Bruno and {Szalay}, Alex S. and {Welsh}, Barry Y. and {Yi}, Sukyoung K.},
        title = "{UV Star Formation Rates in the Local Universe}",
      journal = {\apjs},
         year = 2007,
        month = dec,
       volume = {173},
       number = {2},
        pages = {267-292},
          doi = {10.1086/519218},
archivePrefix = {arXiv},
       eprint = {0704.3611},
 primaryClass = {astro-ph},
       adsurl = {https://ui.adsabs.harvard.edu/abs/2007ApJS..173..267S}
}

@ARTICLE{Magdis2010,
       author = {{Magdis}, G.~E. and {Elbaz}, D. and {Daddi}, E. and {Morrison}, G.~E. and {Dickinson}, M. and {Rigopoulou}, D. and {Gobat}, R. and {Hwang}, H.~S.},
        title = "{A Multi-wavelength View of the Star Formation Activity at z \raisebox{-0.5ex}\textasciitilde 3}",
      journal = {\apj},
         year = 2010,
        month = may,
       volume = {714},
       number = {2},
        pages = {1740-1745},
          doi = {10.1088/0004-637X/714/2/1740},
archivePrefix = {arXiv},
       eprint = {1003.5773},
 primaryClass = {astro-ph.CO},
       adsurl = {https://ui.adsabs.harvard.edu/abs/2010ApJ...714.1740M}
}

@ARTICLE{Rodighiero2011,
       author = {{Rodighiero}, G. and {Daddi}, E. and {Baronchelli}, I. and {Cimatti}, A. and {Renzini}, A. and {Aussel}, H. and {Popesso}, P. and {Lutz}, D. and {Andreani}, P. and {Berta}, S. and {Cava}, A. and {Elbaz}, D. and {Feltre}, A. and {Fontana}, A. and {F{\"o}rster Schreiber}, N.~M. and {Franceschini}, A. and {Genzel}, R. and {Grazian}, A. and {Gruppioni}, C. and {Ilbert}, O. and {Le Floch}, E. and {Magdis}, G. and {Magliocchetti}, M. and {Magnelli}, B. and {Maiolino}, R. and {McCracken}, H. and {Nordon}, R. and {Poglitsch}, A. and {Santini}, P. and {Pozzi}, F. and {Riguccini}, L. and {Tacconi}, L.~J. and {Wuyts}, S. and {Zamorani}, G.},
        title = "{The Lesser Role of Starbursts in Star Formation at z = 2}",
      journal = {\apjl},
         year = 2011,
        month = oct,
       volume = {739},
       number = {2},
          eid = {L40},
        pages = {L40},
          doi = {10.1088/2041-8205/739/2/L40},
archivePrefix = {arXiv},
       eprint = {1108.0933},
 primaryClass = {astro-ph.CO},
       adsurl = {https://ui.adsabs.harvard.edu/abs/2011ApJ...739L..40R}
}

@ARTICLE{Whitaker2014,
       author = {{Whitaker}, Katherine E. and {Franx}, Marijn and {Leja}, Joel and {van Dokkum}, Pieter G. and {Henry}, Alaina and {Skelton}, Rosalind E. and {Fumagalli}, Mattia and {Momcheva}, Ivelina G. and {Brammer}, Gabriel B. and {Labb{\'e}}, Ivo and {Nelson}, Erica J. and {Rigby}, Jane R.},
        title = "{Constraining the Low-mass Slope of the Star Formation Sequence at 0.5 < z < 2.5}",
      journal = {\apj},
         year = 2014,
        month = nov,
       volume = {795},
       number = {2},
          eid = {104},
        pages = {104},
          doi = {10.1088/0004-637X/795/2/104},
archivePrefix = {arXiv},
       eprint = {1407.1843},
 primaryClass = {astro-ph.GA},
       adsurl = {https://ui.adsabs.harvard.edu/abs/2014ApJ...795..104W}
}

@ARTICLE{Heinis2014,
       author = {{Heinis}, S. and {Buat}, V. and {B{\'e}thermin}, M. and {Bock}, J. and {Burgarella}, D. and {Conley}, A. and {Cooray}, A. and {Farrah}, D. and {Ilbert}, O. and {Magdis}, G. and {Marsden}, G. and {Oliver}, S.~J. and {Rigopoulou}, D. and {Roehlly}, Y. and {Schulz}, B. and {Symeonidis}, M. and {Viero}, M. and {Xu}, C.~K. and {Zemcov}, M.},
        title = "{HerMES: dust attenuation and star formation activity in ultraviolet-selected samples from z{\ensuremath{\sim}} 4 to {\ensuremath{\sim}} 1.5}",
      journal = {\mnras},
         year = 2014,
        month = jan,
       volume = {437},
       number = {2},
        pages = {1268-1283},
          doi = {10.1093/mnras/stt1960},
archivePrefix = {arXiv},
       eprint = {1310.3227},
 primaryClass = {astro-ph.CO},
       adsurl = {https://ui.adsabs.harvard.edu/abs/2014MNRAS.437.1268H}
}

@ARTICLE{Tomczak2016,
       author = {{Tomczak}, Adam R. and {Quadri}, Ryan F. and {Tran}, Kim-Vy H. and {Labb{\'e}}, Ivo and {Straatman}, Caroline M.~S. and {Papovich}, Casey and {Glazebrook}, Karl and {Allen}, Rebecca and {Brammer}, Gabreil B. and {Cowley}, Michael and {Dickinson}, Mark and {Elbaz}, David and {Inami}, Hanae and {Kacprzak}, Glenn G. and {Morrison}, Glenn E. and {Nanayakkara}, Themiya and {Persson}, S. Eric and {Rees}, Glen A. and {Salmon}, Brett and {Schreiber}, Corentin and {Spitler}, Lee R. and {Whitaker}, Katherine E.},
        title = "{The SFR-M* Relation and Empirical Star-Formation Histories from ZFOURGE* at 0.5 < z < 4}",
      journal = {\apj},
         year = 2016,
        month = feb,
       volume = {817},
       number = {2},
          eid = {118},
        pages = {118},
          doi = {10.3847/0004-637X/817/2/118},
archivePrefix = {arXiv},
       eprint = {1510.06072},
 primaryClass = {astro-ph.GA},
       adsurl = {https://ui.adsabs.harvard.edu/abs/2016ApJ...817..118T}
}

@ARTICLE{Leslie2020,
       author = {{Leslie}, Sarah K. and {Schinnerer}, Eva and {Liu}, Daizhong and {Magnelli}, Benjamin and {Algera}, Hiddo and {Karim}, Alexander and {Davidzon}, Iary and {Gozaliasl}, Ghassem and {Jim{\'e}nez-Andrade}, Eric F. and {Lang}, Philipp and {Sargent}, Mark T. and {Novak}, Mladen and {Groves}, Brent and {Smol{\v{c}}i{\'c}}, Vernesa and {Zamorani}, Giovanni and {Vaccari}, Mattia and {Battisti}, Andrew and {Vardoulaki}, Eleni and {Peng}, Yingjie and {Kartaltepe}, Jeyhan},
        title = "{The VLA-COSMOS 3 GHz Large Project: Evolution of Specific Star Formation Rates out to z {\ensuremath{\sim}} 5}",
      journal = {\apj},
         year = 2020,
        month = aug,
       volume = {899},
       number = {1},
          eid = {58},
        pages = {58},
          doi = {10.3847/1538-4357/aba044},
archivePrefix = {arXiv},
       eprint = {2006.13937},
 primaryClass = {astro-ph.GA},
       adsurl = {https://ui.adsabs.harvard.edu/abs/2020ApJ...899...58L}
}

@ARTICLE{Leja2022,
       author = {{Leja}, Joel and {Speagle}, Joshua S. and {Ting}, Yuan-Sen and {Johnson}, Benjamin D. and {Conroy}, Charlie and {Whitaker}, Katherine E. and {Nelson}, Erica J. and {Dokkum}, Pieter van and {Franx}, Marijn},
        title = "{A New Census of the 0.2 < z < 3.0 Universe. II. The Star-forming Sequence}",
      journal = {\apj},
         year = 2022,
        month = sep,
       volume = {936},
       number = {2},
          eid = {165},
        pages = {165},
          doi = {10.3847/1538-4357/ac887d},
archivePrefix = {arXiv},
       eprint = {2110.04314},
 primaryClass = {astro-ph.GA},
       adsurl = {https://ui.adsabs.harvard.edu/abs/2022ApJ...936..165L}
}

@ARTICLE{Wuyts2011,
       author = {{Wuyts}, Stijn and {F{\"o}rster Schreiber}, Natascha M. and {van der Wel}, Arjen and {Magnelli}, Benjamin and {Guo}, Yicheng and {Genzel}, Reinhard and {Lutz}, Dieter and {Aussel}, Herv{\'e} and {Barro}, Guillermo and {Berta}, Stefano and {Cava}, Antonio and {Graci{\'a}-Carpio}, Javier and {Hathi}, Nimish P. and {Huang}, Kuang-Han and {Kocevski}, Dale D. and {Koekemoer}, Anton M. and {Lee}, Kyoung-Soo and {Le Floc'h}, Emeric and {McGrath}, Elizabeth J. and {Nordon}, Raanan and {Popesso}, Paola and {Pozzi}, Francesca and {Riguccini}, Laurie and {Rodighiero}, Giulia and {Saintonge}, Amelie and {Tacconi}, Linda},
        title = "{Galaxy Structure and Mode of Star Formation in the SFR-Mass Plane from z \raisebox{-0.5ex}\textasciitilde 2.5 to z \raisebox{-0.5ex}\textasciitilde 0.1}",
      journal = {\apj},
         year = 2011,
        month = dec,
       volume = {742},
       number = {2},
          eid = {96},
        pages = {96},
          doi = {10.1088/0004-637X/742/2/96},
archivePrefix = {arXiv},
       eprint = {1107.0317},
 primaryClass = {astro-ph.CO},
       adsurl = {https://ui.adsabs.harvard.edu/abs/2011ApJ...742...96W}
}

@ARTICLE{Iyer2018,
       author = {{Iyer}, Kartheik and {Gawiser}, Eric and {Dav{\'e}}, Romeel and {Davis}, Philip and {Finkelstein}, Steven L. and {Kodra}, Dritan and {Koekemoer}, Anton M. and {Kurczynski}, Peter and {Newman}, Jeffery A. and {Pacifici}, Camilla and {Somerville}, Rachel S.},
        title = "{The SFR-M $_{*}$ Correlation Extends to Low Mass at High Redshift}",
      journal = {\apj},
         year = 2018,
        month = oct,
       volume = {866},
       number = {2},
          eid = {120},
        pages = {120},
          doi = {10.3847/1538-4357/aae0fa},
archivePrefix = {arXiv},
       eprint = {1809.04099},
 primaryClass = {astro-ph.GA},
       adsurl = {https://ui.adsabs.harvard.edu/abs/2018ApJ...866..120I}
}

@ARTICLE{Oke1983,
       author = {{Oke}, J.~B. and {Gunn}, J.~E.},
        title = "{Secondary standard stars for absolute spectrophotometry.}",
      journal = {\apj},
         year = 1983,
        month = mar,
       volume = {266},
        pages = {713-717},
          doi = {10.1086/160817},
       adsurl = {https://ui.adsabs.harvard.edu/abs/1983ApJ...266..713O}
}

@ARTICLE{Peng2010,
       author = {{Peng}, Ying-jie and {Lilly}, Simon J. and {Kova{\v{c}}}, Katarina and {Bolzonella}, Micol and {Pozzetti}, Lucia and {Renzini}, Alvio and {Zamorani}, Gianni and {Ilbert}, Olivier and {Knobel}, Christian and {Iovino}, Angela and {Maier}, Christian and {Cucciati}, Olga and {Tasca}, Lidia and {Carollo}, C. Marcella and {Silverman}, John and {Kampczyk}, Pawel and {de Ravel}, Loic and {Sanders}, David and {Scoville}, Nicholas and {Contini}, Thierry and {Mainieri}, Vincenzo and {Scodeggio}, Marco and {Kneib}, Jean-Paul and {Le F{\`e}vre}, Olivier and {Bardelli}, Sandro and {Bongiorno}, Angela and {Caputi}, Karina and {Coppa}, Graziano and {de la Torre}, Sylvain and {Franzetti}, Paolo and {Garilli}, Bianca and {Lamareille}, Fabrice and {Le Borgne}, Jean-Francois and {Le Brun}, Vincent and {Mignoli}, Marco and {Perez Montero}, Enrique and {Pello}, Roser and {Ricciardelli}, Elena and {Tanaka}, Masayuki and {Tresse}, Laurence and {Vergani}, Daniela and {Welikala}, Niraj and {Zucca}, Elena and {Oesch}, Pascal and {Abbas}, Ummi and {Barnes}, Luke and {Bordoloi}, Rongmon and {Bottini}, Dario and {Cappi}, Alberto and {Cassata}, Paolo and {Cimatti}, Andrea and {Fumana}, Marco and {Hasinger}, Gunther and {Koekemoer}, Anton and {Leauthaud}, Alexei and {Maccagni}, Dario and {Marinoni}, Christian and {McCracken}, Henry and {Memeo}, Pierdomenico and {Meneux}, Baptiste and {Nair}, Preethi and {Porciani}, Cristiano and {Presotto}, Valentina and {Scaramella}, Roberto},
        title = "{Mass and Environment as Drivers of Galaxy Evolution in SDSS and zCOSMOS and the Origin of the Schechter Function}",
      journal = {\apj},
         year = 2010,
        month = sep,
       volume = {721},
       number = {1},
        pages = {193-221},
          doi = {10.1088/0004-637X/721/1/193},
archivePrefix = {arXiv},
       eprint = {1003.4747},
 primaryClass = {astro-ph.CO},
       adsurl = {https://ui.adsabs.harvard.edu/abs/2010ApJ...721..193P}
}

@ARTICLE{Boquien2019,
       author = {{Boquien}, M. and {Burgarella}, D. and {Roehlly}, Y. and {Buat}, V. and {Ciesla}, L. and {Corre}, D. and {Inoue}, A.~K. and {Salas}, H.},
        title = "{CIGALE: a python Code Investigating GALaxy Emission}",
      journal = {\aap},
         year = 2019,
        month = feb,
       volume = {622},
          eid = {A103},
        pages = {A103},
          doi = {10.1051/0004-6361/201834156},
archivePrefix = {arXiv},
       eprint = {1811.03094},
 primaryClass = {astro-ph.GA},
       adsurl = {https://ui.adsabs.harvard.edu/abs/2019A&A...622A.103B}
}

@ARTICLE{Noll2009,
       author = {{Noll}, S. and {Burgarella}, D. and {Giovannoli}, E. and {Buat}, V. and {Marcillac}, D. and {Mu{\~n}oz-Mateos}, J.~C.},
        title = "{Analysis of galaxy spectral energy distributions from far-UV to far-IR with CIGALE: studying a SINGS test sample}",
      journal = {\aap},
         year = 2009,
        month = dec,
       volume = {507},
       number = {3},
        pages = {1793-1813},
          doi = {10.1051/0004-6361/200912497},
archivePrefix = {arXiv},
       eprint = {0909.5439},
 primaryClass = {astro-ph.CO},
       adsurl = {https://ui.adsabs.harvard.edu/abs/2009A&A...507.1793N}
}

@ARTICLE{Burgarella2005,
       author = {{Burgarella}, D. and {Buat}, V. and {Iglesias-P{\'a}ramo}, J.},
        title = "{Star formation and dust attenuation properties in galaxies from a statistical ultraviolet-to-far-infrared analysis}",
      journal = {\mnras},
         year = 2005,
        month = jul,
       volume = {360},
       number = {4},
        pages = {1413-1425},
          doi = {10.1111/j.1365-2966.2005.09131.x},
archivePrefix = {arXiv},
       eprint = {astro-ph/0504434},
 primaryClass = {astro-ph},
       adsurl = {https://ui.adsabs.harvard.edu/abs/2005MNRAS.360.1413B}
}

@ARTICLE{Shimakawa2015,
       author = {{Shimakawa}, Rhythm and {Kodama}, Tadayuki and {Steidel}, Charles C. and {Tadaki}, Ken-ichi and {Tanaka}, Ichi and {Strom}, Allison L. and {Hayashi}, Masao and {Koyama}, Yusei and {Suzuki}, Tomoko L. and {Yamamoto}, Moegi},
        title = "{Correlation between star formation activity and electron density of ionized gas at z = 2.5}",
      journal = {\mnras},
         year = 2015,
        month = aug,
       volume = {451},
       number = {2},
        pages = {1284-1289},
          doi = {10.1093/mnras/stv915},
archivePrefix = {arXiv},
       eprint = {1411.1408},
 primaryClass = {astro-ph.CO},
       adsurl = {https://ui.adsabs.harvard.edu/abs/2015MNRAS.451.1284S}
}

@article{astropy:2013,
Adsurl = {http://adsabs.harvard.edu/abs/2013A%26A...558A..33A},
Archiveprefix = {arXiv},
Author = {{Astropy Collaboration} and {Robitaille}, T.~P. and {Tollerud}, E.~J. and {Greenfield}, P. and {Droettboom}, M. and {Bray}, E. and {Aldcroft}, T. and {Davis}, M. and {Ginsburg}, A. and {Price-Whelan}, A.~M. and {Kerzendorf}, W.~E. and {Conley}, A. and {Crighton}, N. and {Barbary}, K. and {Muna}, D. and {Ferguson}, H. and {Grollier}, F. and {Parikh}, M.~M. and {Nair}, P.~H. and {Unther}, H.~M. and {Deil}, C. and {Woillez}, J. and {Conseil}, S. and {Kramer}, R. and {Turner}, J.~E.~H. and {Singer}, L. and {Fox}, R. and {Weaver}, B.~A. and {Zabalza}, V. and {Edwards}, Z.~I. and {Azalee Bostroem}, K. and {Burke}, D.~J. and {Casey}, A.~R. and {Crawford}, S.~M. and {Dencheva}, N. and {Ely}, J. and {Jenness}, T. and {Labrie}, K. and {Lim}, P.~L. and {Pierfederici}, F. and {Pontzen}, A. and {Ptak}, A. and {Refsdal}, B. and {Servillat}, M. and {Streicher}, O.},
Doi = {10.1051/0004-6361/201322068},
Eid = {A33},
Eprint = {1307.6212},
Journal = {\aap},
Month = oct,
Pages = {A33},
Primaryclass = {astro-ph.IM},
Title = {{Astropy: A community Python package for astronomy}},
Volume = 558,
Year = 2013}

@ARTICLE{astropy:2018,
       author = {{Astropy Collaboration} and {Price-Whelan}, A.~M. and
         {Sip{\H{o}}cz}, B.~M. and {G{\"u}nther}, H.~M. and {Lim}, P.~L. and
         {Crawford}, S.~M. and {Conseil}, S. and {Shupe}, D.~L. and
         {Craig}, M.~W. and {Dencheva}, N. and {Ginsburg}, A. and {Vand
        erPlas}, J.~T. and {Bradley}, L.~D. and {P{\'e}rez-Su{\'a}rez}, D. and
         {de Val-Borro}, M. and {Aldcroft}, T.~L. and {Cruz}, K.~L. and
         {Robitaille}, T.~P. and {Tollerud}, E.~J. and {Ardelean}, C. and
         {Babej}, T. and {Bach}, Y.~P. and {Bachetti}, M. and {Bakanov}, A.~V. and
         {Bamford}, S.~P. and {Barentsen}, G. and {Barmby}, P. and
         {Baumbach}, A. and {Berry}, K.~L. and {Biscani}, F. and {Boquien}, M. and
         {Bostroem}, K.~A. and {Bouma}, L.~G. and {Brammer}, G.~B. and
         {Bray}, E.~M. and {Breytenbach}, H. and {Buddelmeijer}, H. and
         {Burke}, D.~J. and {Calderone}, G. and {Cano Rodr{\'\i}guez}, J.~L. and
         {Cara}, M. and {Cardoso}, J.~V.~M. and {Cheedella}, S. and {Copin}, Y. and
         {Corrales}, L. and {Crichton}, D. and {D'Avella}, D. and {Deil}, C. and
         {Depagne}, {\'E}. and {Dietrich}, J.~P. and {Donath}, A. and
         {Droettboom}, M. and {Earl}, N. and {Erben}, T. and {Fabbro}, S. and
         {Ferreira}, L.~A. and {Finethy}, T. and {Fox}, R.~T. and
         {Garrison}, L.~H. and {Gibbons}, S.~L.~J. and {Goldstein}, D.~A. and
         {Gommers}, R. and {Greco}, J.~P. and {Greenfield}, P. and
         {Groener}, A.~M. and {Grollier}, F. and {Hagen}, A. and {Hirst}, P. and
         {Homeier}, D. and {Horton}, A.~J. and {Hosseinzadeh}, G. and {Hu}, L. and
         {Hunkeler}, J.~S. and {Ivezi{\'c}}, {\v{Z}}. and {Jain}, A. and
         {Jenness}, T. and {Kanarek}, G. and {Kendrew}, S. and {Kern}, N.~S. and
         {Kerzendorf}, W.~E. and {Khvalko}, A. and {King}, J. and {Kirkby}, D. and
         {Kulkarni}, A.~M. and {Kumar}, A. and {Lee}, A. and {Lenz}, D. and
         {Littlefair}, S.~P. and {Ma}, Z. and {Macleod}, D.~M. and
         {Mastropietro}, M. and {McCully}, C. and {Montagnac}, S. and
         {Morris}, B.~M. and {Mueller}, M. and {Mumford}, S.~J. and {Muna}, D. and
         {Murphy}, N.~A. and {Nelson}, S. and {Nguyen}, G.~H. and
         {Ninan}, J.~P. and {N{\"o}the}, M. and {Ogaz}, S. and {Oh}, S. and
         {Parejko}, J.~K. and {Parley}, N. and {Pascual}, S. and {Patil}, R. and
         {Patil}, A.~A. and {Plunkett}, A.~L. and {Prochaska}, J.~X. and
         {Rastogi}, T. and {Reddy Janga}, V. and {Sabater}, J. and
         {Sakurikar}, P. and {Seifert}, M. and {Sherbert}, L.~E. and
         {Sherwood-Taylor}, H. and {Shih}, A.~Y. and {Sick}, J. and
         {Silbiger}, M.~T. and {Singanamalla}, S. and {Singer}, L.~P. and
         {Sladen}, P.~H. and {Sooley}, K.~A. and {Sornarajah}, S. and
         {Streicher}, O. and {Teuben}, P. and {Thomas}, S.~W. and
         {Tremblay}, G.~R. and {Turner}, J.~E.~H. and {Terr{\'o}n}, V. and
         {van Kerkwijk}, M.~H. and {de la Vega}, A. and {Watkins}, L.~L. and
         {Weaver}, B.~A. and {Whitmore}, J.~B. and {Woillez}, J. and
         {Zabalza}, V. and {Astropy Contributors}},
        title = "{The Astropy Project: Building an Open-science Project and Status of the v2.0 Core Package}",
      journal = {\aj},
         year = 2018,
        month = sep,
       volume = {156},
       number = {3},
          eid = {123},
        pages = {123},
          doi = {10.3847/1538-3881/aabc4f},
archivePrefix = {arXiv},
       eprint = {1801.02634},
 primaryClass = {astro-ph.IM},
       adsurl = {https://ui.adsabs.harvard.edu/abs/2018AJ....156..123A}
}

@ARTICLE{astropy:2022,
       author = {{Astropy Collaboration} and {Price-Whelan}, Adrian M. and {Lim}, Pey Lian and {Earl}, Nicholas and {Starkman}, Nathaniel and {Bradley}, Larry and {Shupe}, David L. and {Patil}, Aarya A. and {Corrales}, Lia and {Brasseur}, C.~E. and {N{"o}the}, Maximilian and {Donath}, Axel and {Tollerud}, Erik and {Morris}, Brett M. and {Ginsburg}, Adam and {Vaher}, Eero and {Weaver}, Benjamin A. and {Tocknell}, James and {Jamieson}, William and {van Kerkwijk}, Marten H. and {Robitaille}, Thomas P. and {Merry}, Bruce and {Bachetti}, Matteo and {G{"u}nther}, H. Moritz and {Aldcroft}, Thomas L. and {Alvarado-Montes}, Jaime A. and {Archibald}, Anne M. and {B{'o}di}, Attila and {Bapat}, Shreyas and {Barentsen}, Geert and {Baz{'a}n}, Juanjo and {Biswas}, Manish and {Boquien}, M{'e}d{'e}ric and {Burke}, D.~J. and {Cara}, Daria and {Cara}, Mihai and {Conroy}, Kyle E. and {Conseil}, Simon and {Craig}, Matthew W. and {Cross}, Robert M. and {Cruz}, Kelle L. and {D'Eugenio}, Francesco and {Dencheva}, Nadia and {Devillepoix}, Hadrien A.~R. and {Dietrich}, J{"o}rg P. and {Eigenbrot}, Arthur Davis and {Erben}, Thomas and {Ferreira}, Leonardo and {Foreman-Mackey}, Daniel and {Fox}, Ryan and {Freij}, Nabil and {Garg}, Suyog and {Geda}, Robel and {Glattly}, Lauren and {Gondhalekar}, Yash and {Gordon}, Karl D. and {Grant}, David and {Greenfield}, Perry and {Groener}, Austen M. and {Guest}, Steve and {Gurovich}, Sebastian and {Handberg}, Rasmus and {Hart}, Akeem and {Hatfield-Dodds}, Zac and {Homeier}, Derek and {Hosseinzadeh}, Griffin and {Jenness}, Tim and {Jones}, Craig K. and {Joseph}, Prajwel and {Kalmbach}, J. Bryce and {Karamehmetoglu}, Emir and {Ka{l}uszy{'n}ski}, Miko{l}aj and {Kelley}, Michael S.~P. and {Kern}, Nicholas and {Kerzendorf}, Wolfgang E. and {Koch}, Eric W. and {Kulumani}, Shankar and {Lee}, Antony and {Ly}, Chun and {Ma}, Zhiyuan and {MacBride}, Conor and {Maljaars}, Jakob M. and {Muna}, Demitri and {Murphy}, N.~A. and {Norman}, Henrik and {O'Steen}, Richard and {Oman}, Kyle A. and {Pacifici}, Camilla and {Pascual}, Sergio and {Pascual-Granado}, J. and {Patil}, Rohit R. and {Perren}, Gabriel I. and {Pickering}, Timothy E. and {Rastogi}, Tanuj and {Roulston}, Benjamin R. and {Ryan}, Daniel F. and {Rykoff}, Eli S. and {Sabater}, Jose and {Sakurikar}, Parikshit and {Salgado}, Jes{'u}s and {Sanghi}, Aniket and {Saunders}, Nicholas and {Savchenko}, Volodymyr and {Schwardt}, Ludwig and {Seifert-Eckert}, Michael and {Shih}, Albert Y. and {Jain}, Anany Shrey and {Shukla}, Gyanendra and {Sick}, Jonathan and {Simpson}, Chris and {Singanamalla}, Sudheesh and {Singer}, Leo P. and {Singhal}, Jaladh and {Sinha}, Manodeep and {Sip{H{o}}cz}, Brigitta M. and {Spitler}, Lee R. and {Stansby}, David and {Streicher}, Ole and {{{S}}umak}, Jani and {Swinbank}, John D. and {Taranu}, Dan S. and {Tewary}, Nikita and {Tremblay}, Grant R. and {Val-Borro}, Miguel de and {Van Kooten}, Samuel J. and {Vasovi{'c}}, Zlatan and {Verma}, Shresth and {de Miranda Cardoso}, Jos{'e} Vin{'i}cius and {Williams}, Peter K.~G. and {Wilson}, Tom J. and {Winkel}, Benjamin and {Wood-Vasey}, W.~M. and {Xue}, Rui and {Yoachim}, Peter and {Zhang}, Chen and {Zonca}, Andrea and {Astropy Project Contributors}},
        title = "{The Astropy Project: Sustaining and Growing a Community-oriented Open-source Project and the Latest Major Release (v5.0) of the Core Package}",
      journal = {apj},
         year = 2022,
        month = aug,
       volume = {935},
       number = {2},
          eid = {167},
        pages = {167},
          doi = {10.3847/1538-4357/ac7c74},
archivePrefix = {arXiv},
       eprint = {2206.14220},
 primaryClass = {astro-ph.IM},
       adsurl = {https://ui.adsabs.harvard.edu/abs/2022ApJ...935..167A}
}

@ARTICLE{Bekki2003,
       author = {{Bekki}, Kenji and {Couch}, Warrick J.},
        title = "{Starbursts from the Strong Compression of Galactic Molecular Clouds due to the High Pressure of the Intracluster Medium}",
      journal = {\apjl},
         year = 2003,
        month = oct,
       volume = {596},
       number = {1},
        pages = {L13-L16},
          doi = {10.1086/379054},
archivePrefix = {arXiv},
       eprint = {astro-ph/0308264},
 primaryClass = {astro-ph},
       adsurl = {https://ui.adsabs.harvard.edu/abs/2003ApJ...596L..13B}
}

@ARTICLE{harris2020array,
 title         = {Array programming with {NumPy}},
 author        = {Charles R. Harris and K. Jarrod Millman and St{\'{e}}fan J.
                 van der Walt and Ralf Gommers and Pauli Virtanen and David
                 Cournapeau and Eric Wieser and Julian Taylor and Sebastian
                 Berg and Nathaniel J. Smith and Robert Kern and Matti Picus
                 and Stephan Hoyer and Marten H. van Kerkwijk and Matthew
                 Brett and Allan Haldane and Jaime Fern{\'{a}}ndez del
                 R{\'{i}}o and Mark Wiebe and Pearu Peterson and Pierre
                 G{\'{e}}rard-Marchant and Kevin Sheppard and Tyler Reddy and
                 Warren Weckesser and Hameer Abbasi and Christoph Gohlke and
                 Travis E. Oliphant},
 year          = {2020},
 month         = sep,
 journal       = {Nature},
 volume        = {585},
 number        = {7825},
 pages         = {357--362},
 doi           = {10.1038/s41586-020-2649-2},
 publisher     = {Springer Science and Business Media {LLC}},
 url           = {https://doi.org/10.1038/s41586-020-2649-2}
}

@software{Photutils,
author       = {Larry Bradley and
                Brigitta Sip{\H o}cz and
                Thomas Robitaille and
                Erik Tollerud and
                Z\`e Vin{\'{\i}}cius and
                Christoph Deil and
                Kyle Barbary and
                Tom J Wilson and
                Ivo Busko and
                Axel Donath and
                Hans Moritz G{\"u}nther and
                Mihai Cara and
                P. L. Lim and
                Sebastian Me{\ss}linger and
                Simon Conseil and
                Azalee Bostroem and
                Michael Droettboom and
                E. M. Bray and
                Lars Andersen Bratholm and
                Geert Barentsen and
                Matt Craig and
                Shivangee Rathi and
                Sergio Pascual and
                Gabriel Perren and
                Iskren Y. Georgiev and
                Miguel de Val-Borro and
                Wolfgang Kerzendorf and
                Yoonsoo P. Bach and
                Bruno Quint and
                Harrison Souchereau},
title        = {astropy/photutils: 1.8.0},
month        = may,
year         = 2023,
publisher    = {Zenodo},
version      = {1.8.0},
doi          = {10.5281/zenodo.7946442},
url          = {https://doi.org/10.5281/zenodo.7946442}
}

@ARTICLE{Jose2023,
       author = {{P{\'e}rez-Mart{\'\i}nez}, J.~M. and {Dannerbauer}, H. and {Kodama}, T. and {Koyama}, Y. and {Shimakawa}, R. and {Suzuki}, T.~L. and {Calvi}, R. and {Chen}, Z. and {Daikuhara}, K. and {Hatch}, N.~A. and {Laza-Ramos}, A. and {Sobral}, D. and {Stott}, J.~P. and {Tanaka}, I.},
        title = "{Signs of environmental effects on star-forming galaxies in the Spiderweb protocluster at z = 2.16}",
      journal = {\mnras},
         year = 2023,
        month = jan,
       volume = {518},
       number = {2},
        pages = {1707-1734},
          doi = {10.1093/mnras/stac2784},
archivePrefix = {arXiv},
       eprint = {2209.13069},
 primaryClass = {astro-ph.GA},
       adsurl = {https://ui.adsabs.harvard.edu/abs/2023MNRAS.518.1707P}
}

@ARTICLE{Tozzi2022,
       author = {{Tozzi}, P. and {Gilli}, R. and {Liu}, A. and {Borgani}, S. and {Lepore}, M. and {Di Mascolo}, L. and {Saro}, A. and {Pentericci}, L. and {Carilli}, C. and {Miley}, G. and {Mroczkowski}, T. and {Pannella}, M. and {Rasia}, E. and {Rosati}, P. and {Anderson}, C.~S. and {Calabr{\'o}}, A. and {Churazov}, E. and {Dannerbauer}, H. and {Feruglio}, C. and {Fiore}, F. and {Gobat}, R. and {Jin}, S. and {Nonino}, M. and {Norman}, C. and {R{\"o}ttgering}, H.~J.~A.},
        title = "{The 700 ks Chandra Spiderweb Field. II. Evidence for inverse-Compton and thermal diffuse emission in the Spiderweb galaxy}",
      journal = {\aap},
         year = 2022,
        month = nov,
       volume = {667},
          eid = {A134},
        pages = {A134},
          doi = {10.1051/0004-6361/202244337},
archivePrefix = {arXiv},
       eprint = {2209.15467},
 primaryClass = {astro-ph.GA},
       adsurl = {https://ui.adsabs.harvard.edu/abs/2022A&A...667A.134T}
}

@ARTICLE{Bunker1995,
       author = {{Bunker}, A.~J. and {Warren}, S.~J. and {Hewett}, P.~C. and {Clements}, D.~L.},
        title = "{On near-infrared H\&alpha searches for high-redshift galaxies}",
      journal = {\mnras},
         year = 1995,
        month = mar,
       volume = {273},
       number = {2},
        pages = {513-516},
          doi = {10.1093/mnras/273.2.513},
archivePrefix = {arXiv},
       eprint = {astro-ph/9501026},
 primaryClass = {astro-ph},
       adsurl = {https://ui.adsabs.harvard.edu/abs/1995MNRAS.273..513B}
}

@ARTICLE{Dannerbauer2014,
       author = {{Dannerbauer}, H. and {Kurk}, J.~D. and {De Breuck}, C. and {Wylezalek}, D. and {Santos}, J.~S. and {Koyama}, Y. and {Seymour}, N. and {Tanaka}, M. and {Hatch}, N. and {Altieri}, B. and {Coia}, D. and {Galametz}, A. and {Kodama}, T. and {Miley}, G. and {R{\"o}ttgering}, H. and {Sanchez-Portal}, M. and {Valtchanov}, I. and {Venemans}, B. and {Ziegler}, B.},
        title = "{An excess of dusty starbursts related to the Spiderweb galaxy}",
      journal = {\aap},
         year = 2014,
        month = oct,
       volume = {570},
          eid = {A55},
        pages = {A55},
          doi = {10.1051/0004-6361/201423771},
archivePrefix = {arXiv},
       eprint = {1410.3730},
 primaryClass = {astro-ph.GA},
       adsurl = {https://ui.adsabs.harvard.edu/abs/2014A&A...570A..55D}
}

@ARTICLE{Naufal2023,
       author = {{Naufal}, Abdurrahman and {Koyama}, Yusei and {Shimakawa}, Rhythm and {Kodama}, Tadayuki},
        title = "{Environmental impacts on the rest-frame UV size and morphology ofstar-forming galaxies at $z\sim2$}",
      journal = {arXiv e-prints},
         year = 2023,
        month = sep,
          eid = {arXiv:2309.15450},
        pages = {arXiv:2309.15450},
          doi = {10.48550/arXiv.2309.15450},
archivePrefix = {arXiv},
       eprint = {2309.15450},
 primaryClass = {astro-ph.GA},
       adsurl = {https://ui.adsabs.harvard.edu/abs/2023arXiv230915450N}
}

@ARTICLE{Jose2024,
       author = {{P{\'e}rez-Mart{\'\i}nez}, J.~M. and {Kodama}, T. and {Koyama}, Y. and {Shimakawa}, R. and {Suzuki}, T.~L. and {Daikuhara}, K. and {Adachi}, K. and {Onodera}, M. and {Tanaka}, I.},
        title = "{Enhanced star formation and metallicity deficit in the USS 1558-003 forming protocluster at z = 2.53}",
      journal = {\mnras},
         year = 2024,
        month = feb,
       volume = {527},
       number = {4},
        pages = {10221-10238},
          doi = {10.1093/mnras/stad3805},
archivePrefix = {arXiv},
       eprint = {2312.03574},
 primaryClass = {astro-ph.GA},
       adsurl = {https://ui.adsabs.harvard.edu/abs/2024MNRAS.52710221P}
}

@ARTICLE{Kurk2000,
       author = {{Kurk}, J.~D. and {R{\"o}ttgering}, H.~J.~A. and {Pentericci}, L. and {Miley}, G.~K. and {van Breugel}, W. and {Carilli}, C.~L. and {Ford}, H. and {Heckman}, T. and {McCarthy}, P. and {Moorwood}, A.},
        title = "{A Search for clusters at high redshift. I. Candidate Lyalpha emitters near 1138-262 at z=2.2}",
      journal = {\aap},
         year = 2000,
        month = jun,
       volume = {358},
        pages = {L1-L4},
          doi = {10.48550/arXiv.astro-ph/0005058},
archivePrefix = {arXiv},
       eprint = {astro-ph/0005058},
 primaryClass = {astro-ph},
       adsurl = {https://ui.adsabs.harvard.edu/abs/2000A&A...358L...1K}
}

@ARTICLE{Roettgering1994,
       author = {{Roettgering}, H.~J.~A. and {Lacy}, M. and {Miley}, G.~K. and {Chambers}, K.~C. and {Saunders}, R.},
        title = "{Samples of ultra-steep spectrum radio sources.}",
      journal = {\aaps},
         year = 1994,
        month = nov,
       volume = {108},
        pages = {79-141},
       adsurl = {https://ui.adsabs.harvard.edu/abs/1994A&AS..108...79R}
}

@ARTICLE{Pentericci1997,
       author = {{Pentericci}, L. and {Roettgering}, H.~J.~A. and {Miley}, G.~K. and {Carilli}, C.~L. and {McCarthy}, P.},
        title = "{The radio galaxy 1138-262 at z=2.2: a giant elliptical galaxy at the center of a proto-cluster?}",
      journal = {\aap},
         year = 1997,
        month = oct,
       volume = {326},
        pages = {580-596},
       adsurl = {https://ui.adsabs.harvard.edu/abs/1997A&A...326..580P}
}

@ARTICLE{DiMascolo2023,
       author = {{Di Mascolo}, Luca and {Saro}, Alexandro and {Mroczkowski}, Tony and {Borgani}, Stefano and {Churazov}, Eugene and {Rasia}, Elena and {Tozzi}, Paolo and {Dannerbauer}, Helmut and {Basu}, Kaustuv and {Carilli}, Christopher L. and {Ginolfi}, Michele and {Miley}, George and {Nonino}, Mario and {Pannella}, Maurilio and {Pentericci}, Laura and {Rizzo}, Francesca},
        title = "{Forming intracluster gas in a galaxy protocluster at a redshift of 2.16}",
      journal = {\nat},
         year = 2023,
        month = mar,
       volume = {615},
       number = {7954},
        pages = {809-812},
          doi = {10.1038/s41586-023-05761-x},
archivePrefix = {arXiv},
       eprint = {2303.16226},
 primaryClass = {astro-ph.CO},
       adsurl = {https://ui.adsabs.harvard.edu/abs/2023Natur.615..809D}
}

@ARTICLE{Hatch2011,
       author = {{Hatch}, N.~A. and {Kurk}, J.~D. and {Pentericci}, L. and {Venemans}, B.~P. and {Kuiper}, E. and {Miley}, G.~K. and {R{\"o}ttgering}, H.~J.~A.},
        title = "{H{\ensuremath{\alpha}} emitters in z{\ensuremath{\sim}} 2 protoclusters: evidence for faster evolution in dense environments}",
      journal = {\mnras},
         year = 2011,
        month = aug,
       volume = {415},
       number = {4},
        pages = {2993-3005},
          doi = {10.1111/j.1365-2966.2011.18735.x},
archivePrefix = {arXiv},
       eprint = {1103.4364},
 primaryClass = {astro-ph.CO},
       adsurl = {https://ui.adsabs.harvard.edu/abs/2011MNRAS.415.2993H}
}

@ARTICLE{Kodama2007,
       author = {{Kodama}, Tadayuki and {Tanaka}, Ichi and {Kajisawa}, Masaru and {Kurk}, Jaron and {Venemans}, Bram and {De Breuck}, Carlos and {Vernet}, Jo{\"e}l and {Lidman}, Chris},
        title = "{The first appearance of the red sequence of galaxies in proto-clusters at 2 <\raisebox{-0.5ex}\textasciitilde z <\raisebox{-0.5ex}\textasciitilde 3}",
      journal = {\mnras},
         year = 2007,
        month = jun,
       volume = {377},
       number = {4},
        pages = {1717-1725},
          doi = {10.1111/j.1365-2966.2007.11739.x},
archivePrefix = {arXiv},
       eprint = {astro-ph/0703382},
 primaryClass = {astro-ph},
       adsurl = {https://ui.adsabs.harvard.edu/abs/2007MNRAS.377.1717K}
}

@ARTICLE{Zirm2008,
       author = {{Zirm}, Andrew W. and {Stanford}, S.~A. and {Postman}, M. and {Overzier}, R.~A. and {Blakeslee}, J.~P. and {Rosati}, P. and {Kurk}, J. and {Pentericci}, L. and {Venemans}, B. and {Miley}, G.~K. and {R{\"o}ttgering}, H.~J.~A. and {Franx}, M. and {van der Wel}, A. and {Demarco}, R. and {van Breugel}, W.},
        title = "{The Nascent Red Sequence at z \raisebox{-0.5ex}\textasciitilde 2}",
      journal = {\apj},
         year = 2008,
        month = jun,
       volume = {680},
       number = {1},
        pages = {224-231},
          doi = {10.1086/587449},
archivePrefix = {arXiv},
       eprint = {0802.2095},
 primaryClass = {astro-ph},
       adsurl = {https://ui.adsabs.harvard.edu/abs/2008ApJ...680..224Z}
}

@ARTICLE{Tanaka2010,
       author = {{Tanaka}, M. and {De Breuck}, C. and {Venemans}, B. and {Kurk}, J.},
        title = "{The environmental dependence of galaxy properties at z = 2}",
      journal = {\aap},
         year = 2010,
        month = jul,
       volume = {518},
          eid = {A18},
        pages = {A18},
          doi = {10.1051/0004-6361/200913939},
archivePrefix = {arXiv},
       eprint = {1005.2253},
 primaryClass = {astro-ph.CO},
       adsurl = {https://ui.adsabs.harvard.edu/abs/2010A&A...518A..18T}
}

@BOOK{Osterbrock2006,
       author = {{Osterbrock}, Donald E. and {Ferland}, Gary J.},
        title = "{Astrophysics of gaseous nebulae and active galactic nuclei}",
         year = 2006,
       adsurl = {https://ui.adsabs.harvard.edu/abs/2006agna.book.....O}
}




\appendix
\section{Comparison with JWST Data}

We compare SFR and stellar mass with those obtained from MAHALO deep survey conducted \ha\ NB (NB2071) imaging using Subaru/MOIRCS \citep{Shimakawa2018pks,Daikuhara2024}.
From this dataset, we compare SFRs based on H$\alpha$ emission and stellar mass derived from SED fitting using consists of Subaru / S-Cam / B and z$^{\prime}$ \citep[][]{koyama2013}, HST / ACS / F814W and F475W \citep[][]{Miley2006}, VLT / HAWK-I / Y, H, and Ks \citep[][]{Miley2006,Dannerbauer2014}, Subaru / MOIRCS / J, and Ks \citep[][]{koyama2013}, Subaru / MOIRCS / NB2071 \citep[][]{koyama2013,Shimakawa2018pks}, and Spitzer / IRAC / 3.6 $\mu$m, 4.5 $\mu$m \citep[][]{Seymour2007}. 
For consistency with this study, we re-performed the SED fitting following the methodology adopted in this paper. 
We note that this modification did not alter the global trend of the MS. 
The purpose of this modification is to reduce systematic uncertainties and to demonstrate the difference between HAEs and PBEs.

As described in the main text, SFR and stellar mass of HAEs and PBEs are a $\sim 70\%$ overlap as shown Figure~\ref{fig:comp}. 
The remaining fraction is affected by the influence of surrounding objects or uncertainty of photometric error. 
We emphasize that while the detection of PBEs with small EWs is challenging, PBEs can still be regarded as a useful tracer of star-forming galaxies at the high-$z$ Universe. 

\begin{figure*}
  \begin{tabular}{cc}
    \begin{minipage}{0.50\textwidth}
      \begin{center}
        \includegraphics[width=\columnwidth]{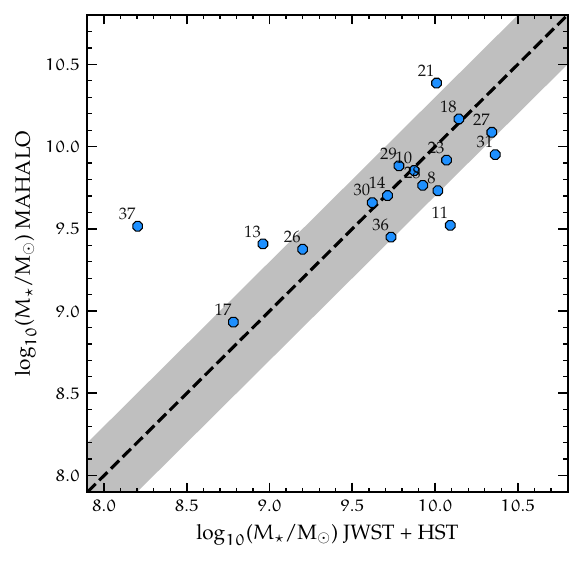}
      \end{center}
    \end{minipage}
    \begin{minipage}{0.50\textwidth}
      \begin{center}
        \includegraphics[width=\columnwidth]{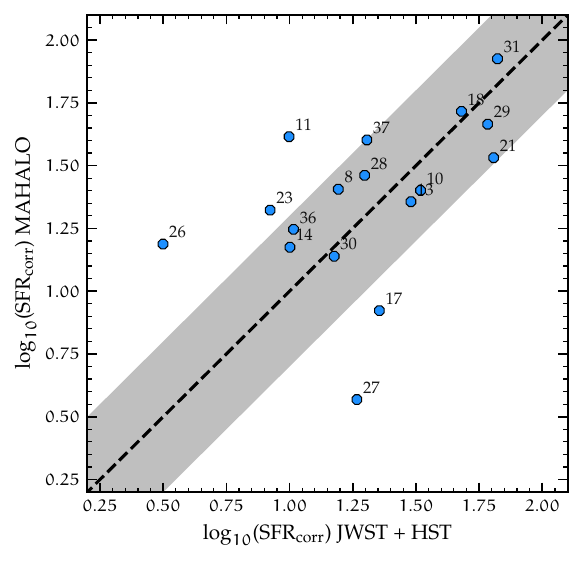}
      \end{center}
    \end{minipage}
  \end{tabular}
\caption{A comparison of physical properties derived from JWST+HST (horizontal axis) and from the MAHALO dataset (vertical axis). 
The left panel shows stellar masses, while the right panel presents star formation rates (SFRs) corrected for dust attenuation. 
In both panels, individual sources are plotted, and the one-to-one relation is drawn as a reference line. 
The shaded regions indicate $\pm0.3$~dex offsets from the one-to-one relation, within which approximately $70\%$ of the sources are located. 
This indicates that the SED fitting based on JWST+HST photometry yields results broadly consistent with those obtained from the MAHALO dataset, without altering the overall main-sequence trend.
}
\label{fig:comp}
\end{figure*}

\section{Cutout images of Spiderweb \pab\ emitters}\label{sec:B}

Figure~\ref{fig:cuts} -- Figure~\ref{fig:cute} show cutout images of Spiderweb PBEs. 
From left to right: JWST/NIRCam F115W, F182M, F410M, the continuum image used for line subtraction, 
the Pa$\beta$ line image, and the ground-based H$\alpha$ narrow-band image from the MAHALO dataset \citep{Daikuhara2024}. 
All panels are $3\arcsec\times3\arcsec$; JWST images adopt a pixel scale of $0\farcs060$~pixel$^{-1}$, 
while the H$\alpha$ panel adopts $0\farcs117$~pixel$^{-1}$ for MOIRCS. 
Blue contours in the H$\alpha$ panel indicate $5\sigma$, $7\sigma$, and $10\sigma$ isophotes, where $\sigma$ is the background RMS 
estimated from a $10\arcsec\times10\arcsec$ box around the target after $2\sigma$ clipping. 

\begin{figure*}
\includegraphics[width=2\columnwidth]{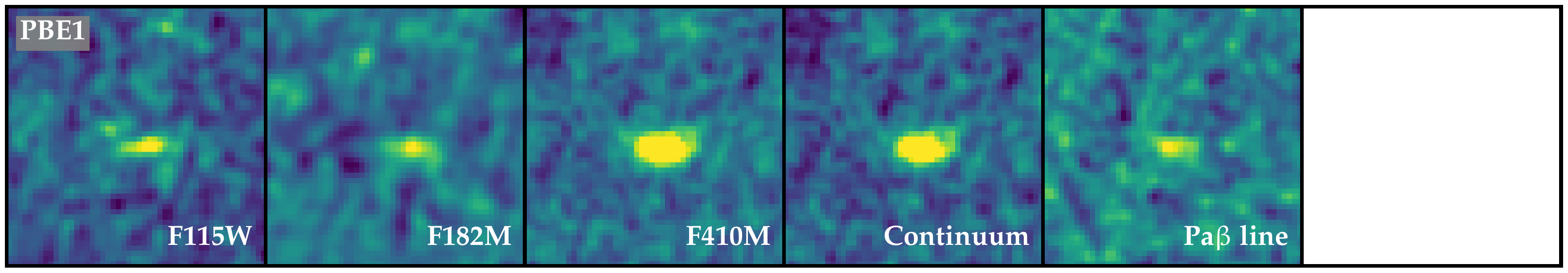}
\includegraphics[width=2\columnwidth]{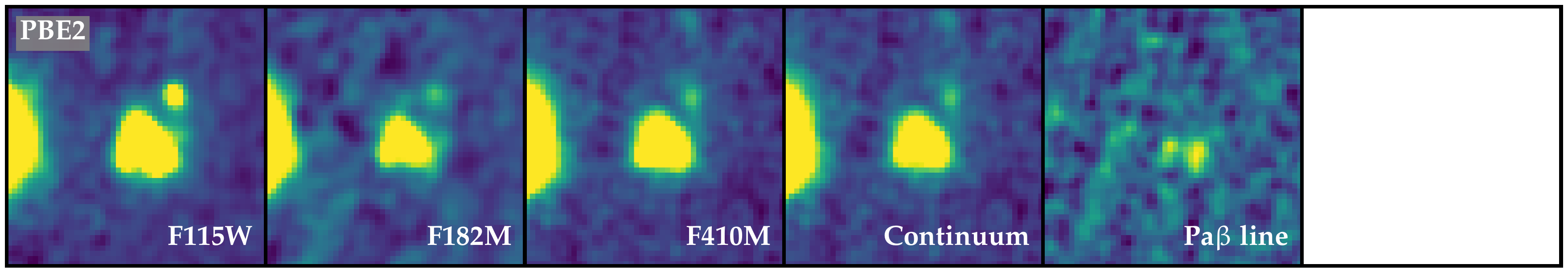}
\includegraphics[width=2\columnwidth]{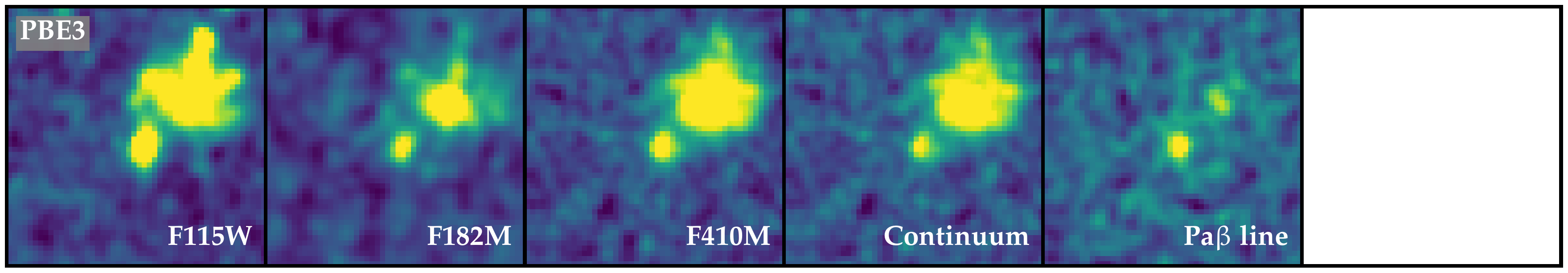}
\includegraphics[width=2\columnwidth]{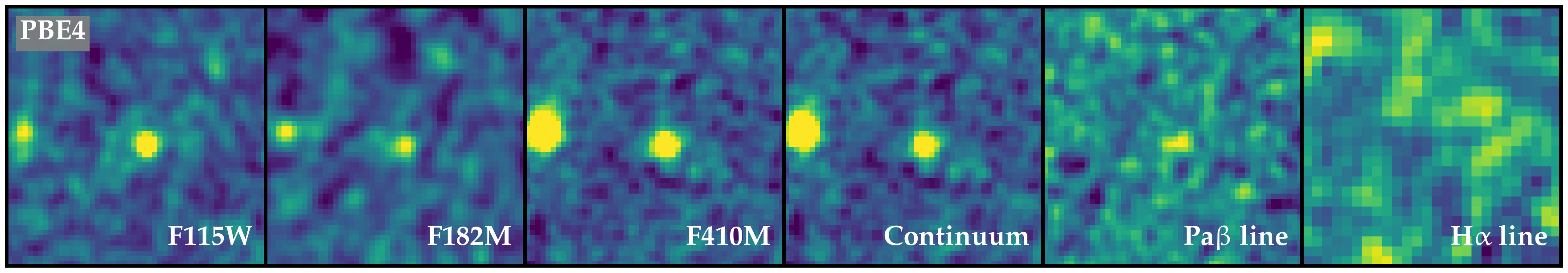}
\includegraphics[width=2\columnwidth]{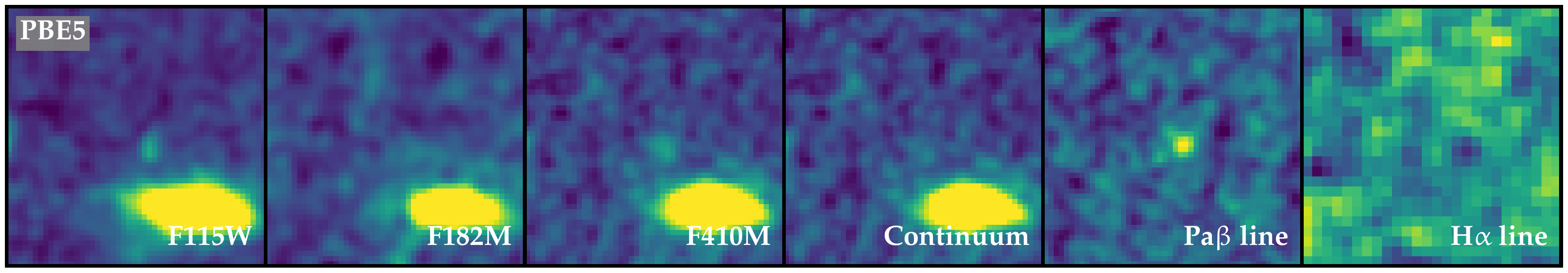}
\includegraphics[width=2\columnwidth]{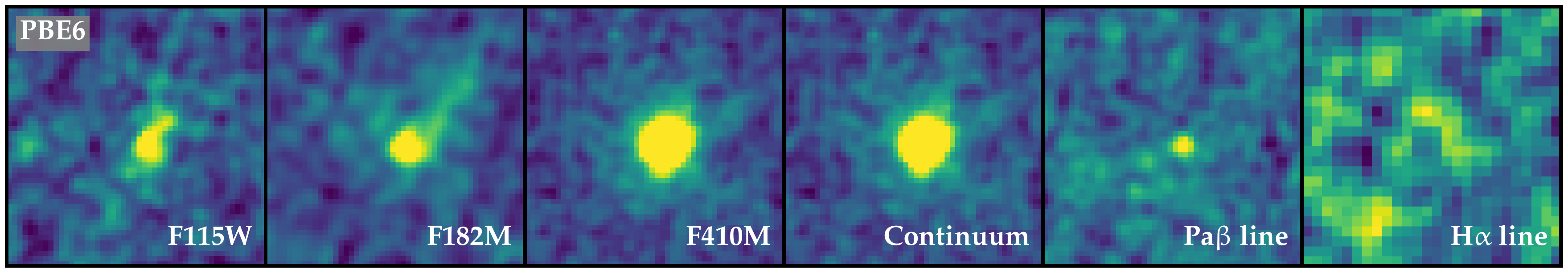}
\includegraphics[width=2\columnwidth]{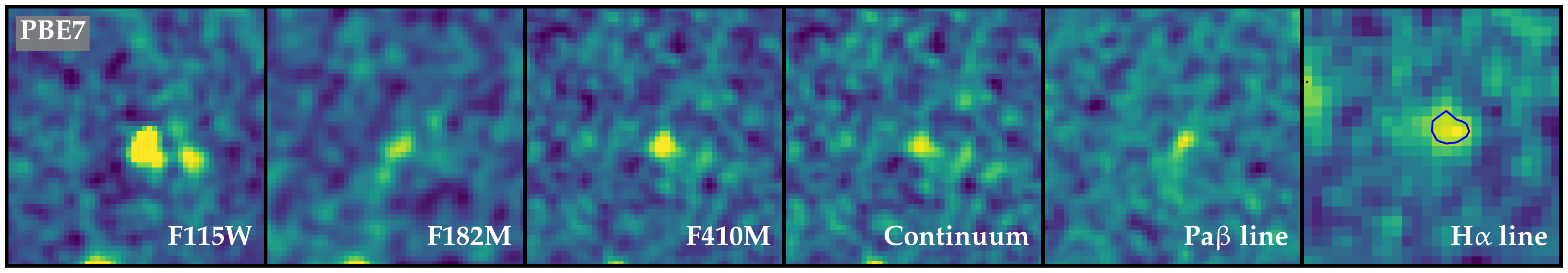}
\caption{The F115W, F182M, F410M, continuum \pab\ and \ha\ cutout images ($3\arcsec\times3\arcsec$) for PBEs.
Contours of \ha\ cutout image are drawn at $5\sigma$, $7\sigma$, and $10\sigma$ above the local background, where $\sigma$ is the $2\sigma$-clipped RMS measured in a $10\arcsec\times10\arcsec$ box centered on each target. }
\label{fig:cuts}
\end{figure*}

\begin{figure*}
\includegraphics[width=2\columnwidth]{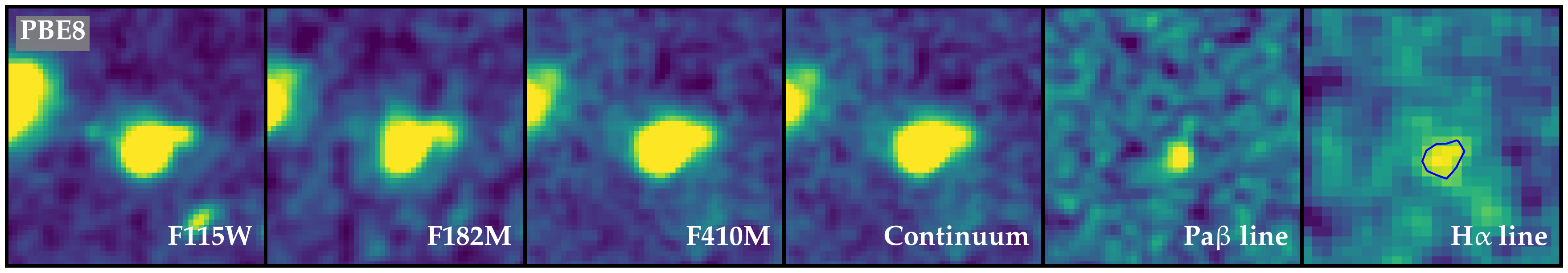}
\includegraphics[width=2\columnwidth]{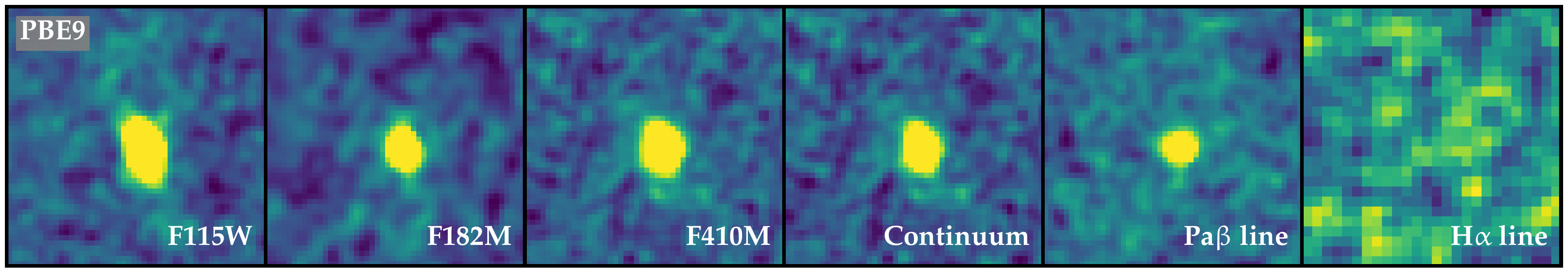}
\includegraphics[width=2\columnwidth]{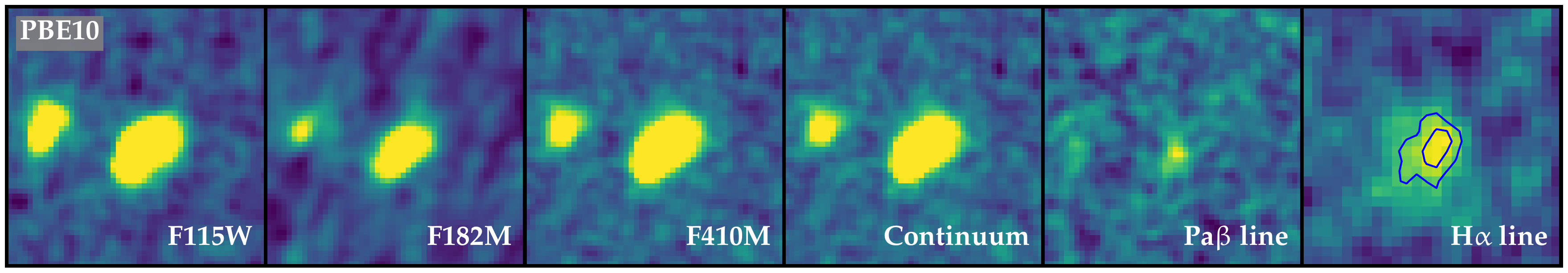}
\includegraphics[width=2\columnwidth]{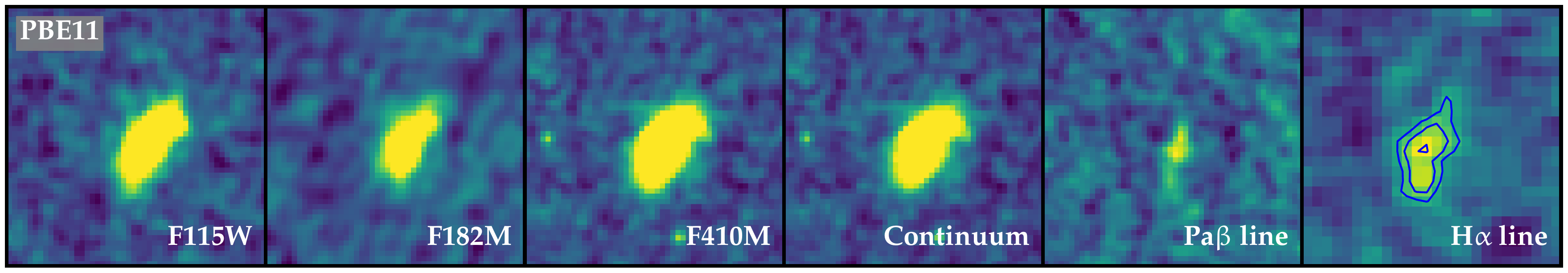}
\includegraphics[width=2\columnwidth]{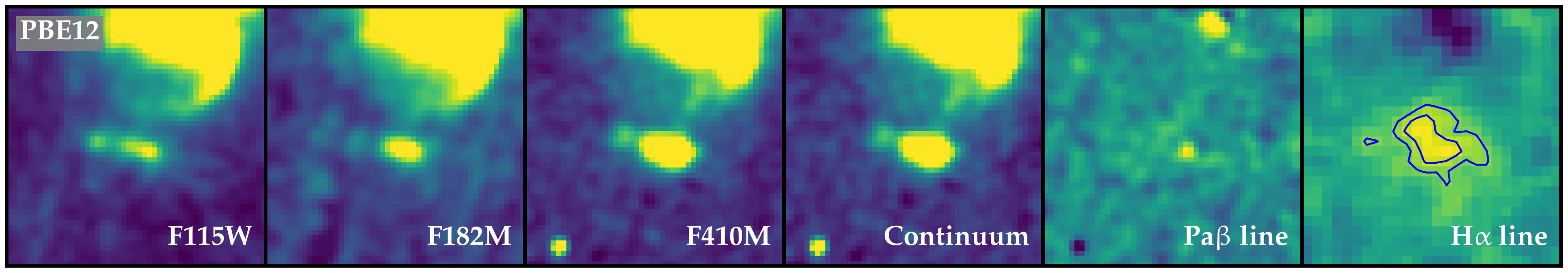}
\includegraphics[width=2\columnwidth]{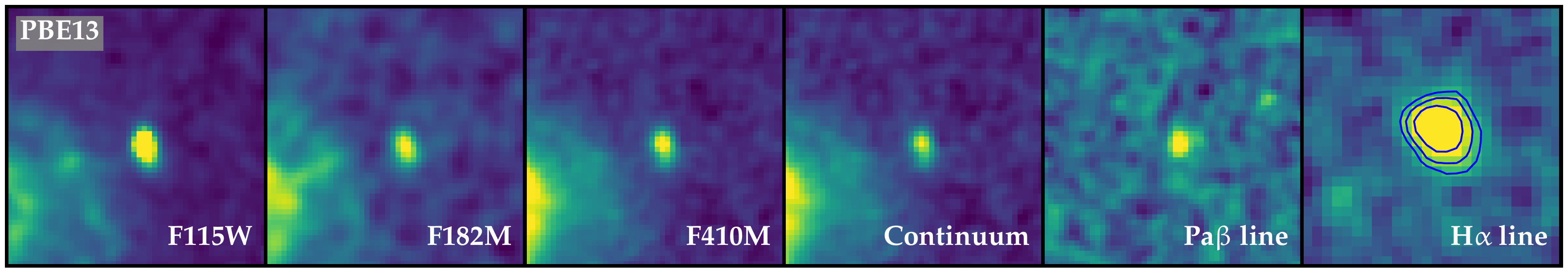}
\includegraphics[width=2\columnwidth]{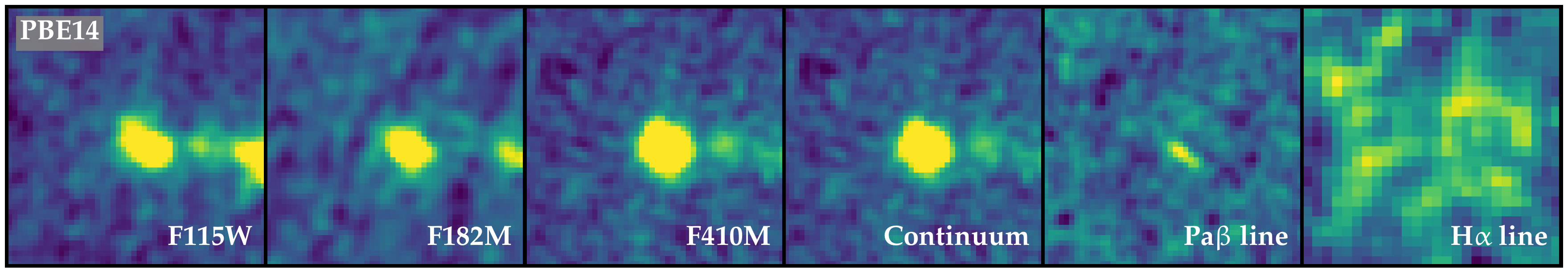}
\caption{(Continued)}
\end{figure*}

\begin{figure*}
\includegraphics[width=2\columnwidth]{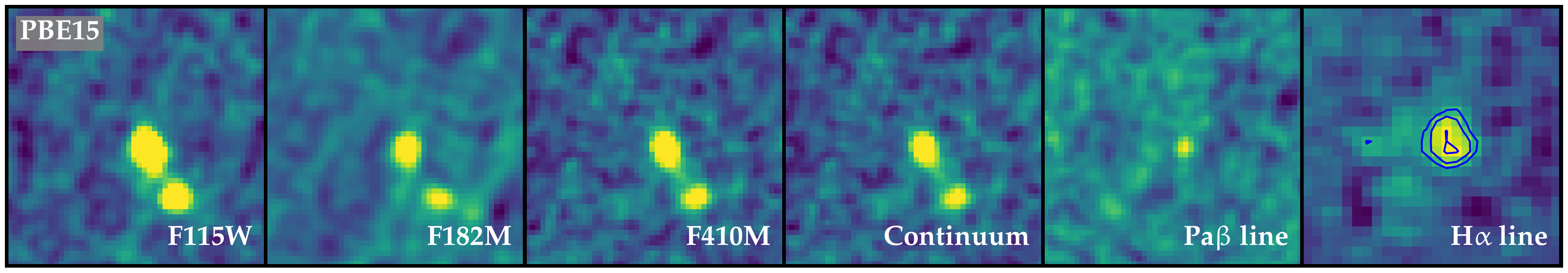}
\includegraphics[width=2\columnwidth]{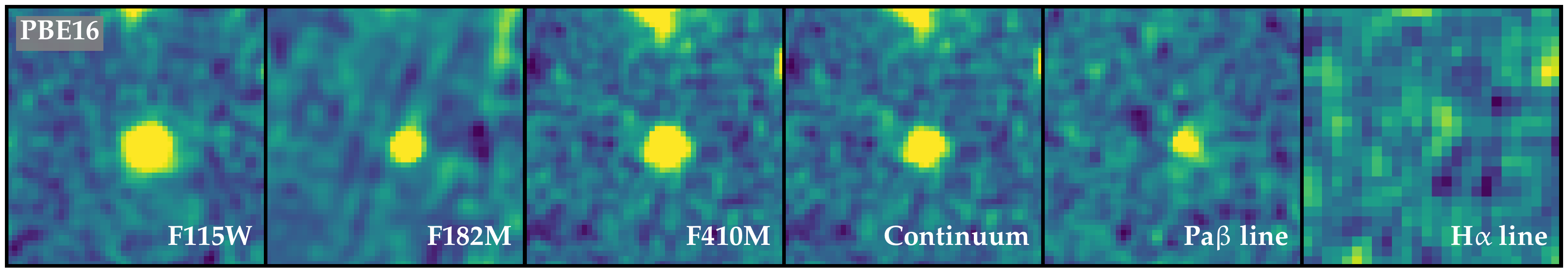}
\includegraphics[width=2\columnwidth]{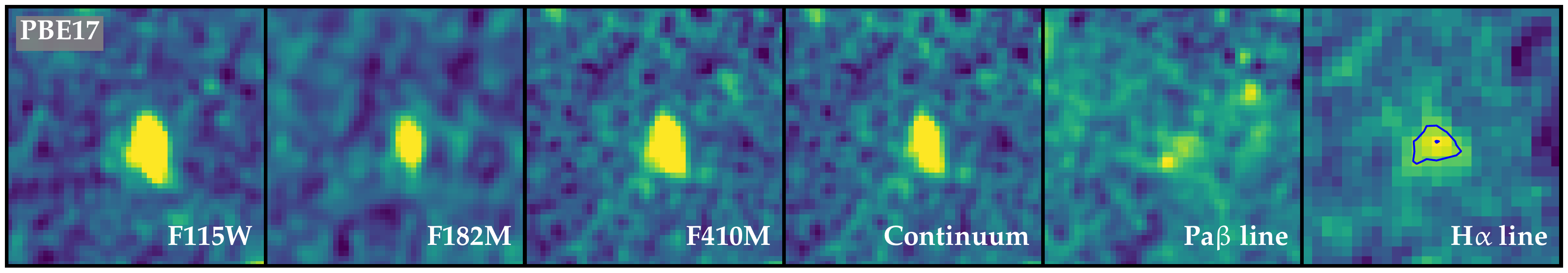}
\includegraphics[width=2\columnwidth]{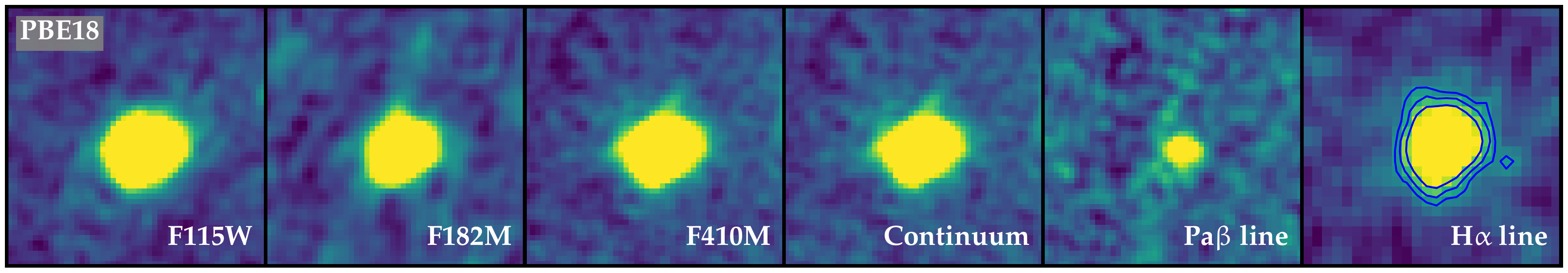}
\includegraphics[width=2\columnwidth]{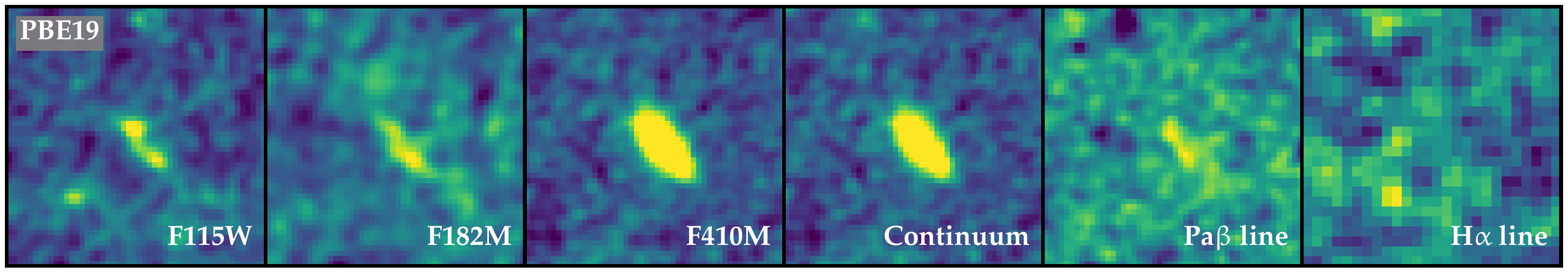}
\includegraphics[width=2\columnwidth]{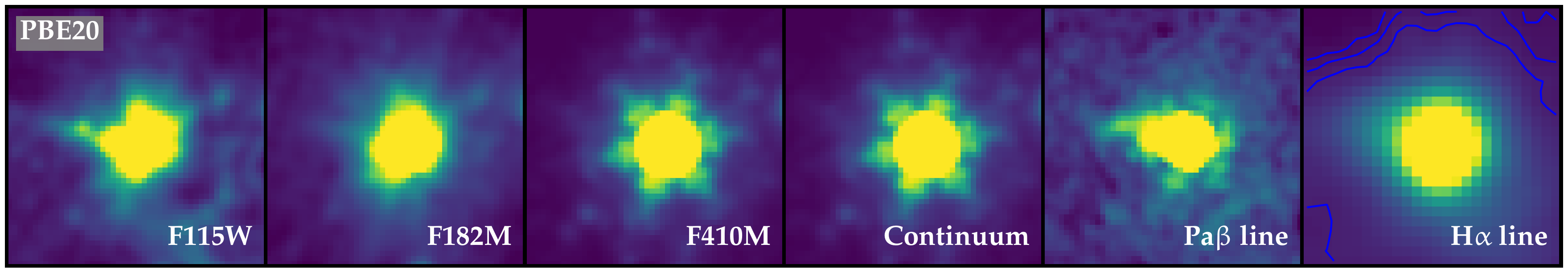}
\includegraphics[width=2\columnwidth]{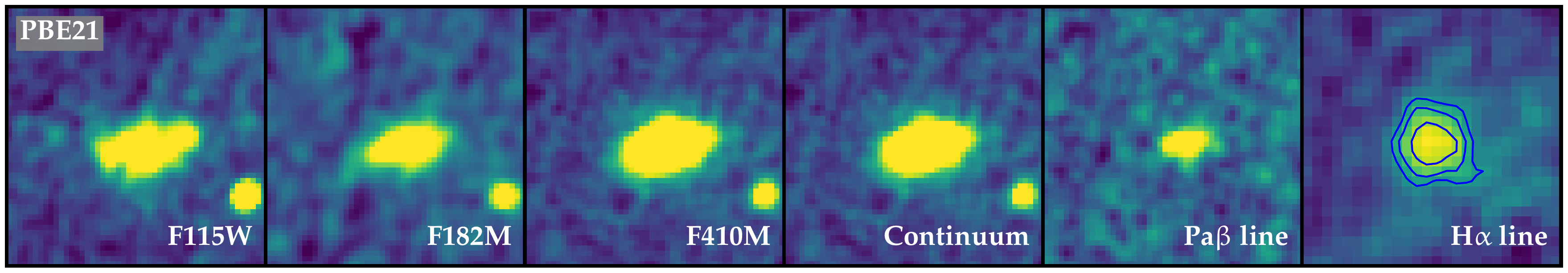}
\caption{(Continued)}
\end{figure*}

\begin{figure*}
\includegraphics[width=2\columnwidth]{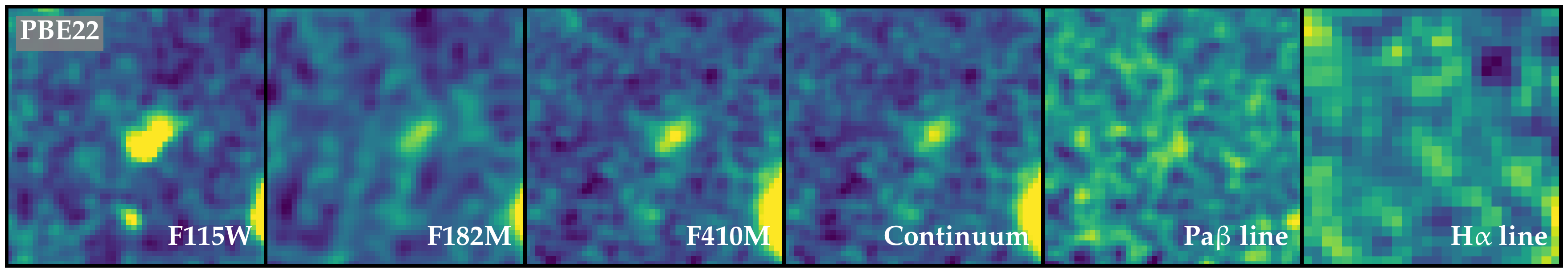}
\includegraphics[width=2\columnwidth]{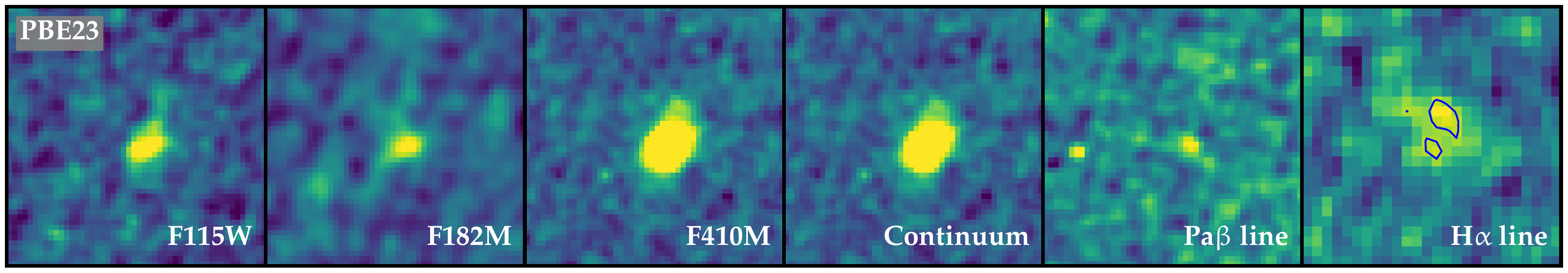}
\includegraphics[width=2\columnwidth]{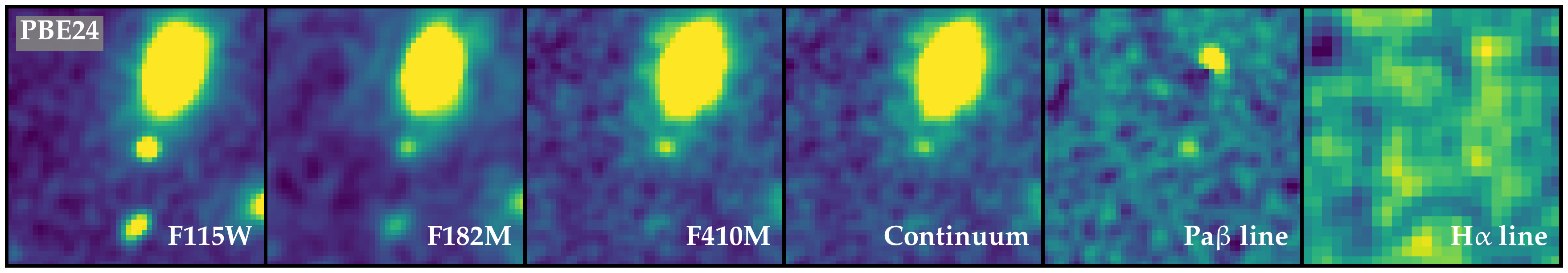}
\includegraphics[width=2\columnwidth]{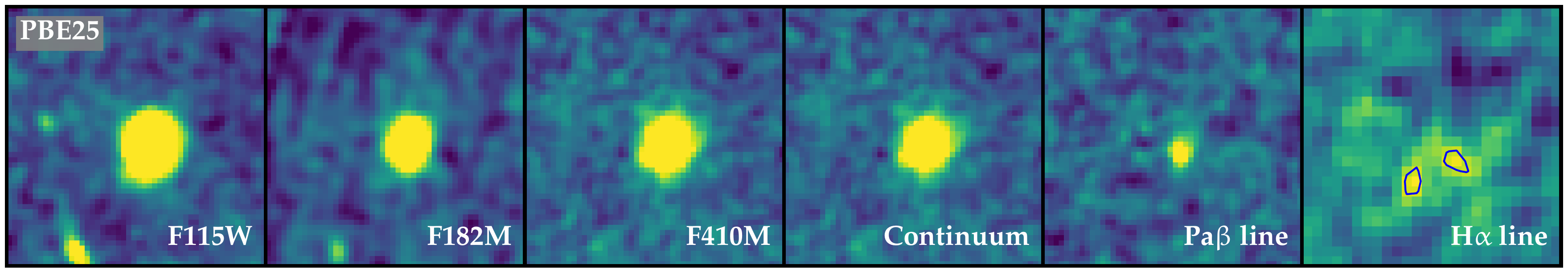}
\includegraphics[width=2\columnwidth]{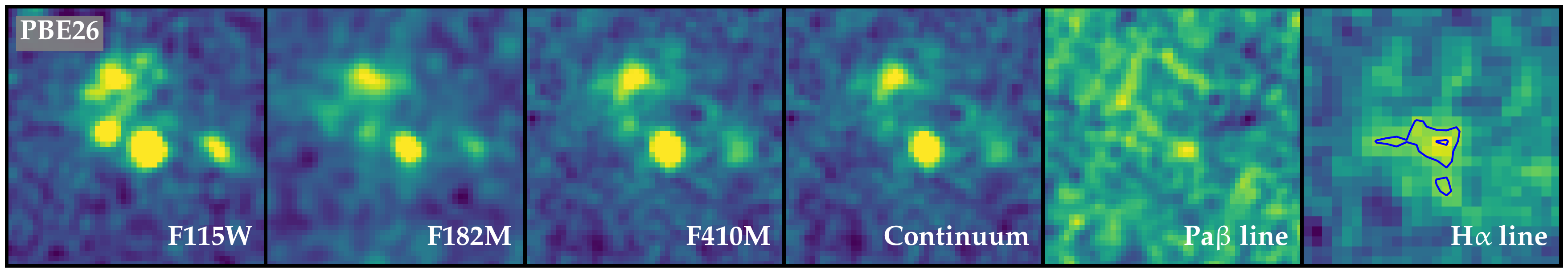}
\includegraphics[width=2\columnwidth]{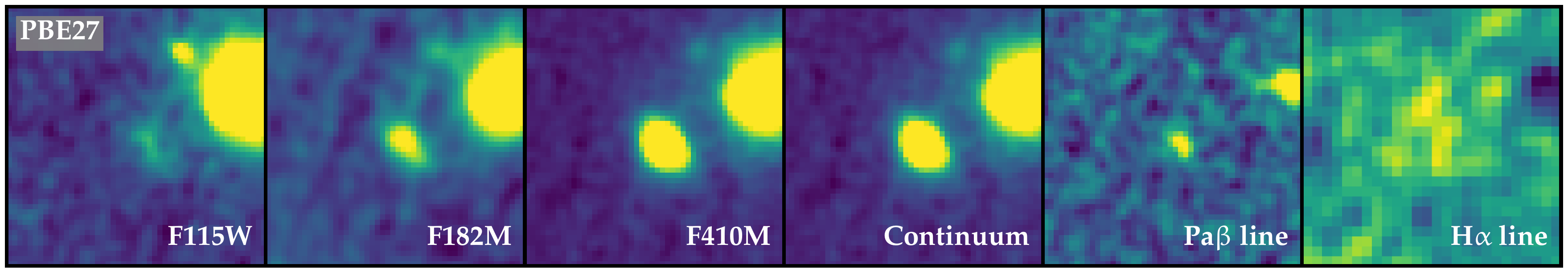}
\includegraphics[width=2\columnwidth]{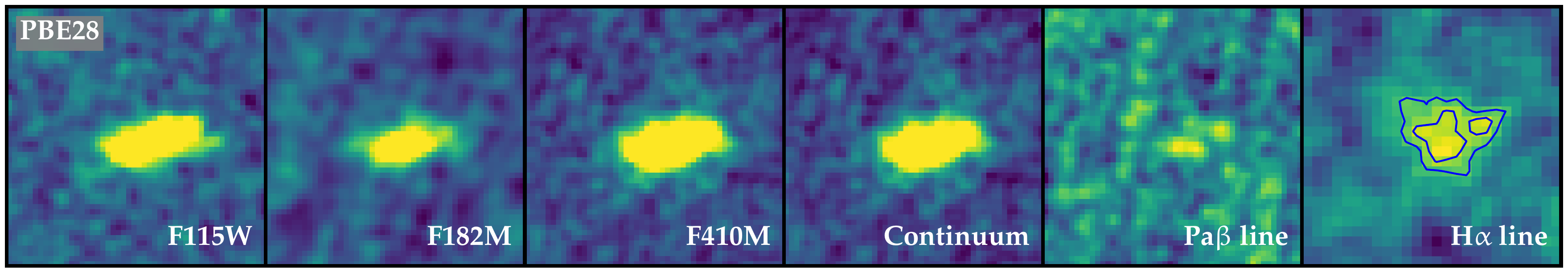}
\caption{(Continued)}
\end{figure*}

\begin{figure*}
\includegraphics[width=2\columnwidth]{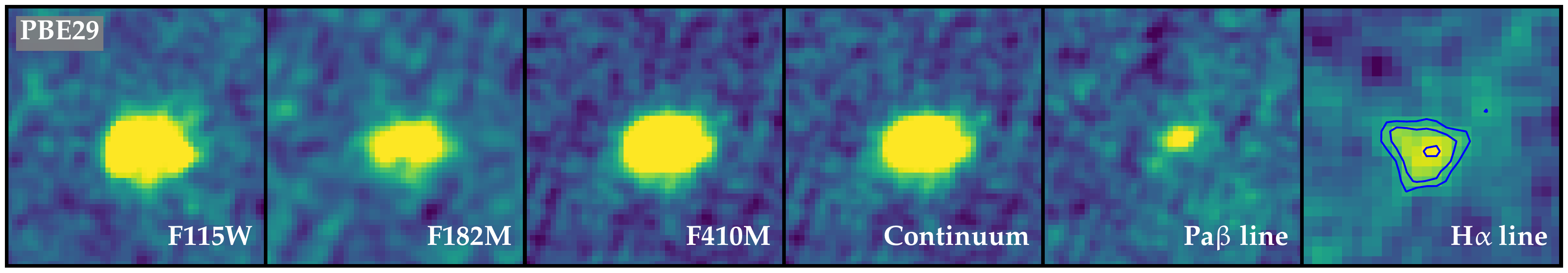}
\includegraphics[width=2\columnwidth]{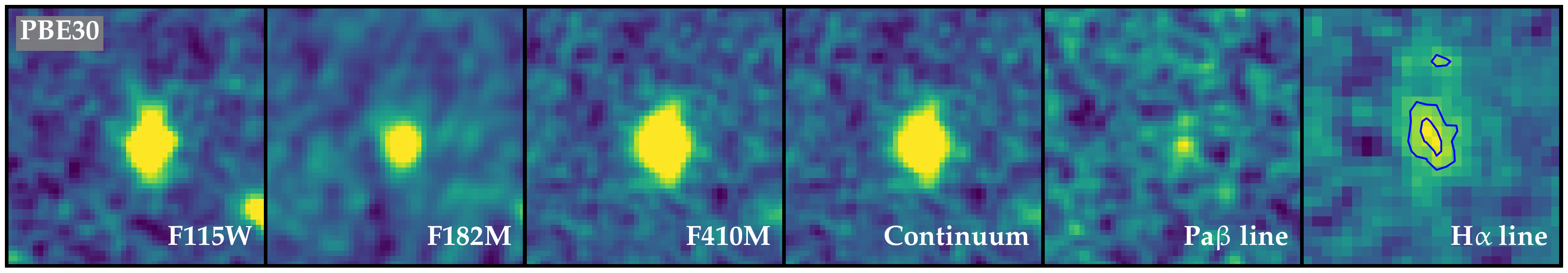}
\includegraphics[width=2\columnwidth]{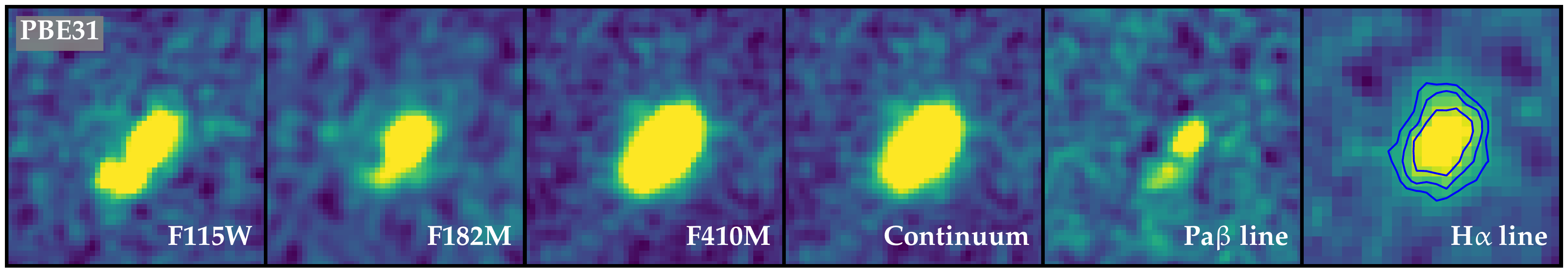}
\includegraphics[width=2\columnwidth]{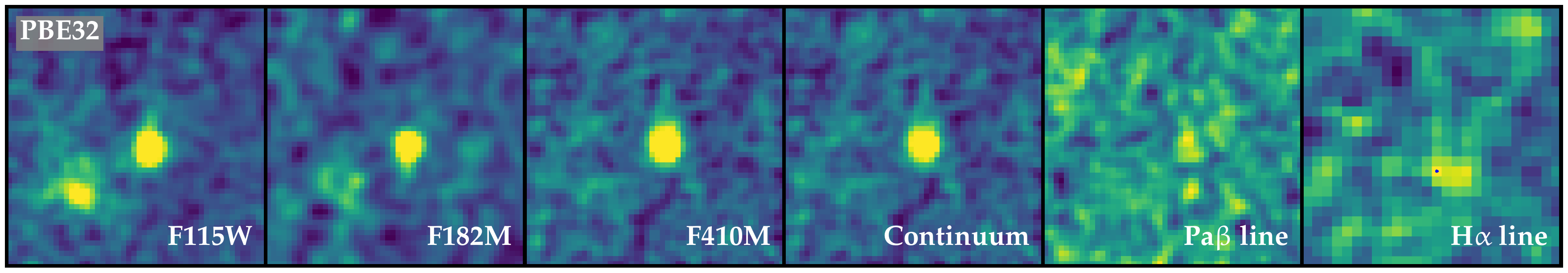}
\includegraphics[width=2\columnwidth]{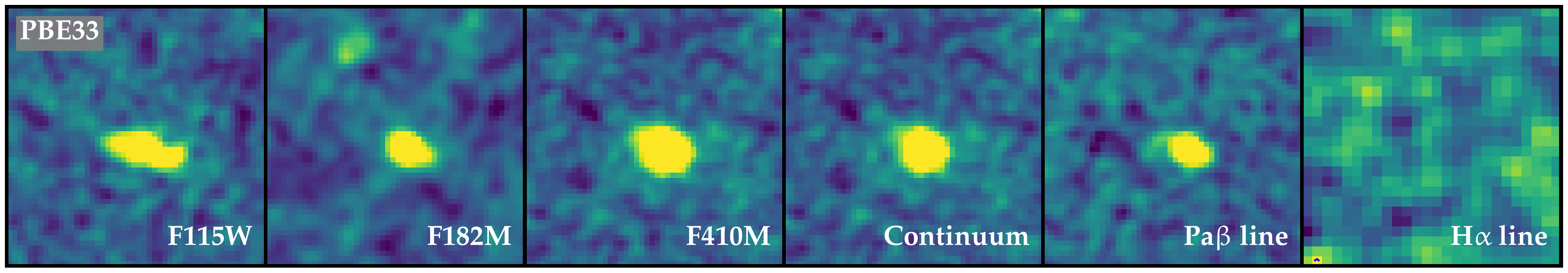}
\includegraphics[width=2\columnwidth]{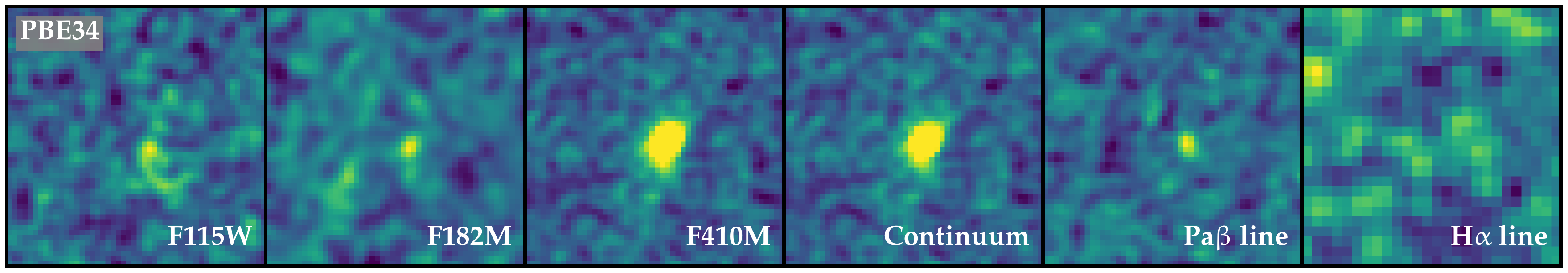}
\includegraphics[width=2\columnwidth]{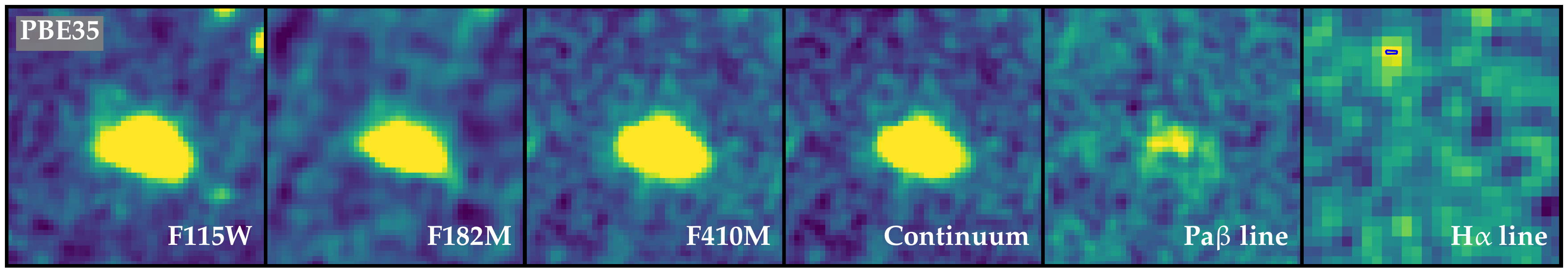}
\caption{(Continued)}
\end{figure*}

\begin{figure*}
\includegraphics[width=2\columnwidth]{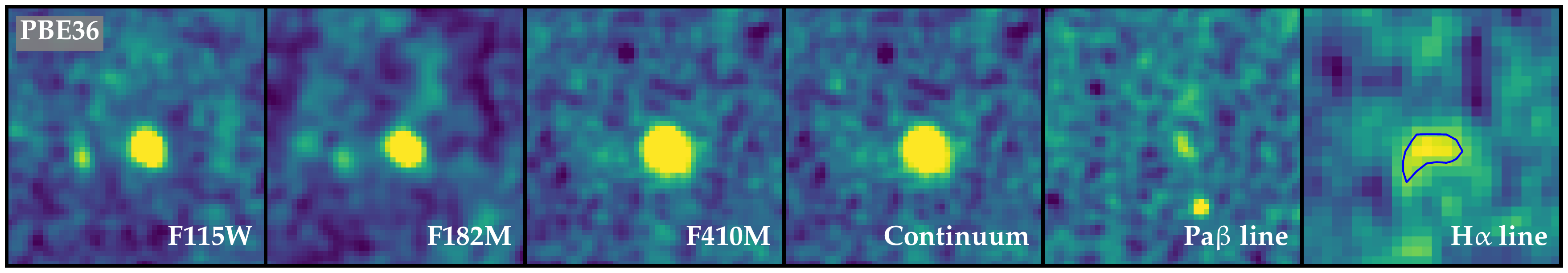}
\includegraphics[width=2\columnwidth]{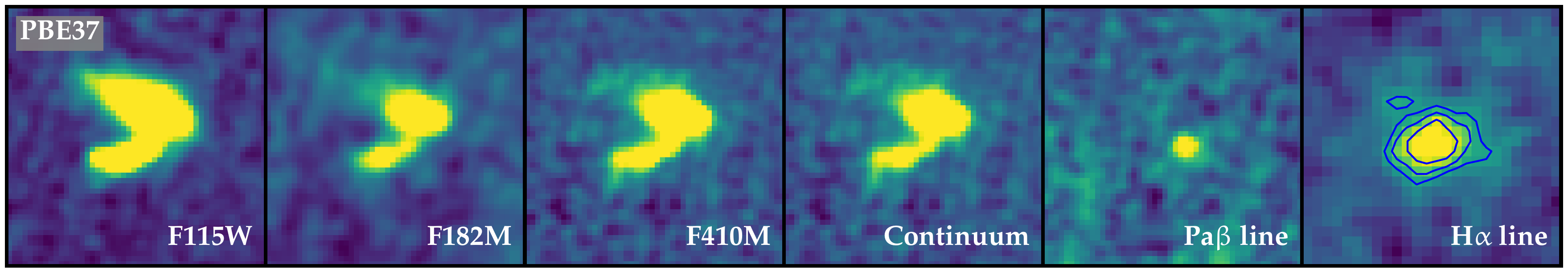}
\includegraphics[width=2\columnwidth]{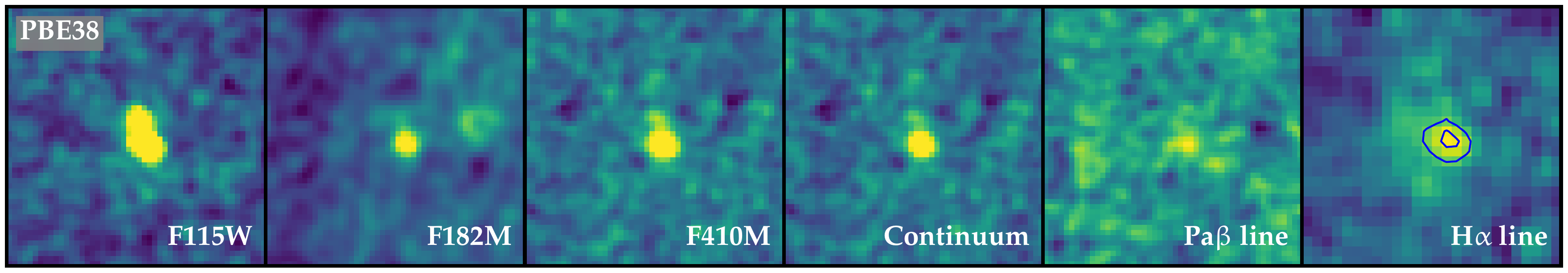}
\includegraphics[width=2\columnwidth]{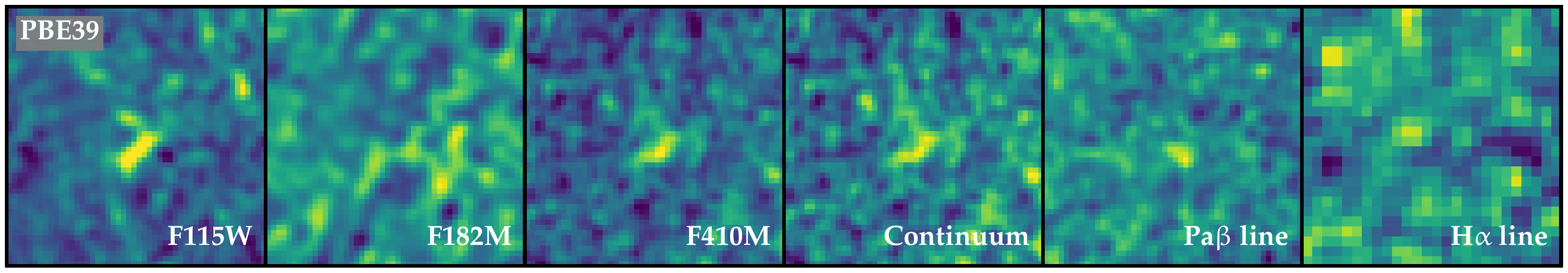}
\includegraphics[width=2\columnwidth]{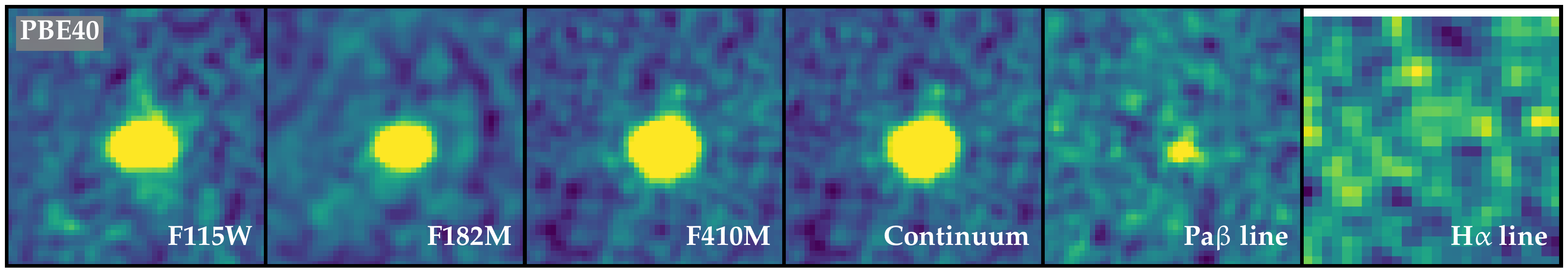}
\includegraphics[width=2\columnwidth]{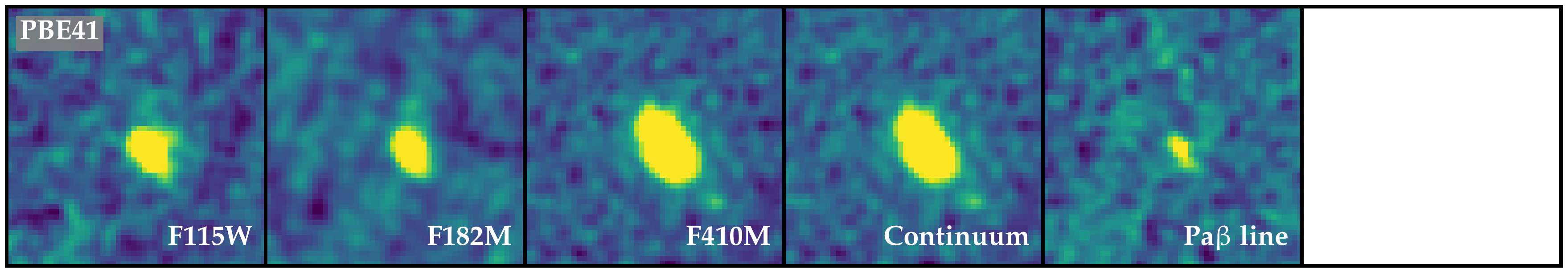}
\caption{(Continued)}
\label{fig:cute}
\end{figure*}


\bsp	
\label{lastpage}
\end{document}